\documentclass[12pt,epsf]{article}
 \usepackage[compat=1.1.0]{tikz-feynman}
 \usepackage{verbatim,graphics,graphicx,color,slashed,amsmath}
 \usepackage{bm}
 \usepackage{cite}
 \usepackage{amssymb}
 \usepackage{braket}
 \usepackage{multirow}
\usepackage[]{hyperref}%my macbook pro
 \hypersetup{colorlinks,bookmarks,unicode,linktocpage=true,linkcolor=blue, anchorcolor=blue, citecolor=blue}
\usepackage{comment}
\usepackage{float}
\newcounter{num}

\begin{document}
\thispagestyle{empty}
\vspace*{-15mm}
%----------
\baselineskip 1pt
\begin{flushright}
\begin{tabular}{l}
%{\bf December 2025}
YITP-25-189, WUCG-25-15
\end{tabular}
\end{flushright}
\baselineskip 24pt
\vglue 10mm

%%%%%%%%%%%%%%%%%%%%%%%%%%%%%%%%%%%%%%%%%%%%%%
%                Title
%%%%%%%%%%%%%%%%%%%%%%%%%%%%%%%%%%%%%%%%%%%%%%
\vspace{15mm}
\begin{center}
{\Large\bf Higgs-Portal Stueckelberg Dark Matter}
\vspace{7mm}

\baselineskip 18pt
{\bf Antonio De Felice${}^{1,\, *}$, Takehiro Ogura${}^{2,\, \ddagger}$, \\Shinji Tsujikawa${}^{3,\, \S}$, Kimiko Yamashita${}^{2,\, \P}$}
\vspace{2mm}

{\it \small
${}^{1}$Center for Gravitational Physics and Quantum Information, Yukawa Institute for Theoretical Physics, Kyoto University, Kyoto 606-8502, Japan\\
${}^{2}$Department of Physics, Ibaraki University, Mito 310-8512, Japan\\
${}^{3}$Department of Physics, Waseda University, 3-4-1 Okubo, Shinjuku, Tokyo 169-8555, Japan
\newline \newline
${}^{*}$antonio.defelice@yukawa.kyoto-u.ac.jp,
${}^{\ddagger}$25nm014a@vc.ibaraki.ac.jp,\\
${}^{\S}$tsujikawa@waseda.jp,
${}^{\P}$kimiko.yamashita.nd93@vc.ibaraki.ac.jp}\\
%\vspace{10mm}
\end{center}
%%%%%%%%%%%%%%%%%%%%%%%%%%%%%%%%%%%%%%%%
%%%%%                              %%%%%
%%%%%          Abstract            %%%%%
%%%%%                              %%%%%
%%%%%%%%%%%%%%%%%%%%%%%%%%%%%%%%%%%%%%%%
\begin{center}
\begin{minipage}{14cm}
\baselineskip 16pt
\noindent
%%%%%---------------------------------

%%%%%--------------------------------
\end{minipage}
\end{center}
\begin{abstract}
We study dark photon dark matter $X$ associated with a dark $U(1)_X$ gauge symmetry.
To evade laboratory and cosmological constraints on kinetic mixing with the Standard Model $U(1)_Y$, 
we assign a $Z_2$-odd dark parity to $X$ that forbids such mixing.
The leading interactions then arise from gauge-invariant dimension-6 Higgs-portal operators, 
including both parity-even and parity-odd terms.
We assume that the dark matter mass is generated via the Stueckelberg mechanism.
In the interacting Stueckelberg EFT, a gauge-invariant dimension-6 parent
reduces in unitary gauge to the nominal dimension-4 Higgs-portal interaction
$(H^\dagger H)X_\mu X^\mu$, with its coefficient correlated with
$m_X^2/\Lambda^2$, while additional derivative operators are organized by the
same longitudinal-mode power counting.
We investigate freeze-in production of $X$ from Higgs-pair annihilations after reheating, 
incorporating both field-strength and Stueckelberg-allowed operators.
First, we consider the case in which the Wilson coefficients of the gauge-invariant dimension-6 operators, $C$ and $\tilde{C}$, are of order unity.
Under a conservative single-cutoff power counting in which the strong-coupling scales of the
longitudinal mode lie above the cutoff of the field-strength effective theory, we find that the
Stueckelberg contributions remain subdominant in dark matter production.
We emphasize that this ordering of scales is an explicit assumption rather than a universal
consequence of perturbative unitarity.
Next, we explore a more general situation in which $C$ and $\tilde{C}$ are smaller than unity.
In this second case, Stueckelberg-allowed operators can become comparable 
and generate contributions proportional to $T_{\rm reh}^3$, $T_{\rm reh}^5$, and
$T_{\rm reh}^7$, including sign-dependent interference effects.
For both cases, we show that there exists a wide parameter space consistent with the observed dark matter relic density.
\end{abstract}

%%%----------------------------------
%%%
%%%
%%%   Main body of the paper
%%%
%%%
%%%----------------------------------
\baselineskip 18pt
\def\thefootnote{\fnsymbol{footnote}}
\setcounter{footnote}{0}
%%%%%%%%%%%

%%%%%%%%%%%%%
\newpage

\section{Introduction}\label{sec:introduction}

Dark photons, the gauge bosons of an additional $U(1)_X$ symmetry, 
have been extensively studied as mediators between the dark sector and the Standard Model (SM)~\cite{Holdom:1985ag} 
(see~\cite{Fabbrichesi:2020wbt} for a review).  
In the conventional setup, the dark photon interacts with the SM particles through kinetic mixing 
with the $U(1)_Y$ field strength. However, current laboratory and cosmological constraints severely restrict the allowed kinetic mixing parameter across a wide mass range~\cite{Caputo:2021eaa}.  
For TeV-scale dark photons, collider constraints lead to a quartic suppression of the small mixing angle in the cross section of $pp \to X \to \ell \ell$ at the 13~TeV Large Hadron Collider, yielding values well below the order of fb for a mixing angle of 0.01~\cite{Yamashita:2024krp}.

Motivated by these limitations on tiny or relatively small mixing angles, our previous work (by one of the authors, K.~Y.)~\cite{Yamashita:2024krp} 
considered a scenario in which kinetic mixing is forbidden by imposing 
a dark $Z_2$ parity under which the dark photon is odd and all SM fields are even.  
In this setup, the leading interactions arise from gauge-invariant dimension-6 Higgs-portal operators, namely  
$(H^\dagger H)X_{\mu\nu} X^{\mu\nu}$ and $(H^\dagger H)\tilde{X}_{\mu\nu}X^{\mu\nu}$,  
allowing the dark photon to be stable and serve as a viable dark matter (DM) candidate.  
In that work, K.~Y. studied the thermal freeze-out production mechanism and identified 
the parameter space consistent with the observed DM relic density and direct-detection bounds.  
In the limit of vanishing momentum transfer, the parity-odd operator is irrelevant for direct detection,  
whereas the parity-even operator is strongly constrained.  
Consequently, the activation of the parity-odd operator is essential to achieve the observed DM relic density.

In this work, we examine the freeze-in production of DM and extend 
the previous scenario by adopting the Stueckelberg mechanism 
as the origin of the dark photon mass.  
The Stueckelberg mass generation preserves the ``$U(1)_X$ gauge symmetry"%
\footnote{The vector field remains invariant under the gauge symmetry. 
Here, double quotation marks are used to indicate the ``fake" $U(1)_X$ gauge symmetry~\cite{Kribs:2022gri}.}.  
Without introducing a dark scalar to generate the mass (as in the Higgs mechanism), the mass term is instead realized through the Stueckelberg framework, which remains valid only at energies below the scale at which perturbative unitarity is violated.
A crucial consequence is that a gauge-invariant Stueckelberg parent reduces in
unitary gauge to the nominal dimension-4 interaction
$(H^\dagger H)\,X_\mu X^\mu$\footnote{The thermal relic abundance of a weakly interacting massive particle
arising from this Stueckelberg-allowed unitary-gauge interaction is discussed in
Ref.~\cite{Lebedev:2011iq} by Lebedev, Lee, and Mambrini.}. Its coefficient is
not independent, but is correlated with $m_X^2/\Lambda^2$. Additional
derivative operators relevant to Higgs-pair annihilation into DM pairs are also
permitted, on top of the field-strength operators considered previously.
We investigate the freeze-in production of Stueckelberg dark photon DM through the Higgs portal, incorporating the dimension-4 and dimension-6 interactions retained in our EFT benchmark.
In the interacting EFT considered here, formulating the mass sector with the
gauge-invariant Stueckelberg building block does more than simply rewrite the
free Proca mass. It correlates the nominal dimension-4 portal coupling with $m_X^2/\Lambda^2$,
introduces separate strong-coupling scales for the longitudinal polarization,
and organizes derivative operators whose freeze-in yields have different
powers of $T_{\rm reh}$.
Consequently, the mass-generation sector can leave a characteristic imprint through
$T_{\rm reh}^5$ and $T_{\rm reh}^7$ terms and through interference between the
Stueckelberg-allowed operators. We identify the assumptions under which these effects are
subleading and show that they can become phenomenologically relevant when the coefficients
of the gauge-invariant dimension-6 operators are suppressed relative to those of the
Stueckelberg-allowed operators.

The paper is organized as follows.  
In Section~\ref{sec:model}, we review the dimension-6 Higgs-portal dark photon model of Ref.~\cite{Yamashita:2024krp} and introduce its Stueckelberg extension.  
In Section~\ref{sec:dm_relic}, we compute the DM relic density via Higgs-portal freeze-in, taking into account the gauge-invariant parity-even and parity-odd dimension-6 terms, and identify the viable parameter space.  
We also discuss the additional operators arising from the Stueckelberg mechanism and compute the corresponding DM relic density, focusing on the parameter region in which their contributions become relevant alongside the gauge-invariant dimension-6 operators.
We then discuss the cosmological scope of the parameter space, including structure-formation
and reheating-related qualifications.
A derivation of the parity-odd Stueckelberg contribution and quantitative
checks of the freeze-in, thermal-mass, and structure-formation estimates are
given in the appendixes.
A summary is presented in Sec.~\ref{sec:summary}.

\section{Higgs-Portal Dark Photon with Stueckelberg Mass}\label{sec:model}
In this section, we first briefly review the Higgs-portal dark photon DM model up to dimension-6 operators, following Ref.~\cite{Yamashita:2024krp}.
We then introduce the Stueckelberg mechanism for generating the dark photon mass without explicitly introducing a dark scalar.

\subsection{Higgs-Portal Interactions at Dimension-6}\label{sec:eft}
Here, we briefly review the dark photon DM model discussed in Ref.~\cite{Yamashita:2024krp}.  
We consider a dark photon associated with an additional dark gauge symmetry $U(1)_X$, 
where the dark photon is odd under a dark parity, whereas all SM fields are neutral under $U(1)_X$ 
and even under the dark parity.  
In this setup, kinetic mixing is absent, and the dark photon can serve as a viable DM candidate.

The gauge-invariant interactions between the dark photon and the SM particles up to dimension~6 are described by the Higgs-portal operators:
\begin{align}
\mathcal{O} &= (H^\dagger H)\, X_{\mu\nu} X^{\mu\nu}, \label{eq:op1} \\
\tilde{\mathcal{O}} &= (H^\dagger H)\, \tilde{X}_{\mu\nu} X^{\mu\nu}, \label{eq:op2}
\end{align}
where $H$ is the SM Higgs doublet,  
$X_{\mu\nu} = \partial_\mu X_\nu - \partial_\nu X_\mu$ is the field strength of the dark photon DM $X_\mu$, and  
$\tilde{X}_{\mu\nu} = (1/2) \epsilon_{\mu\nu\rho\sigma} X^{\rho\sigma}$ is its dual, with $\epsilon_{0123} = 1$.  
The operator $\tilde{\mathcal{O}}$ is odd under parity and time-reversal transformations due to the presence of the Levi-Civita tensor $\epsilon_{\mu\nu\rho\sigma} $.

The effective Lagrangian describing a dark photon with mass $m_X$ and its interactions with the SM particles up to dimension-6 is given by
\begin{align}
\mathcal{L}_\mathrm{VDM}
= \frac{m_X^2}{2}\, X_\mu X^\mu
 + \frac{C}{\Lambda^2}\, \mathcal{O}
 + \frac{\tilde{C}}{\Lambda^2}\, \tilde{\mathcal{O}}, \label{eq:lag}
\end{align}
where $C$ and $\tilde{C}$ are the Wilson coefficients associated 
with the operators in Eqs.~\eqref{eq:op1} and \eqref{eq:op2}, and $\Lambda$ denotes the characteristic energy scale of new physics.
 
We employ the Stueckelberg mechanism to generate the dark photon mass, as described in the following Section~\ref{sec:stueckelberg}.

%%%%%%%%%%%%%
\subsection[Higgs-Portal Stueckelberg Dark Photon DM up to Dimension-6]{Higgs-Portal Stueckelberg Dark Photon DM\\ up to \mbox{Dimension-6}}\label{sec:stueckelberg}
To generate the mass of the dark photon without introducing an explicit dark scalar, we employ the Stueckelberg mechanism (see Ref.~\cite{Ruegg:2003ps} for a review).  
In this setup, the dark photon acquires a mass while preserving the ``$U(1)_X$ gauge symmetry", and the dark parity $X^\mu \to -X^\mu$ ensures the stability of the dark photon as a DM candidate.  

In the Stueckelberg mechanism, we define a ``fake" $U(1)_X$ gauge field~\cite{Kribs:2022gri}:
\begin{align}
A^{\mu} \equiv X^\mu +\frac{\partial^\mu\pi}{m_X}, \label{eq:X}
\end{align}
where $\pi$ is a Stueckelberg scalar field.
Both fields transform under this ``U(1) gauge transformation" as
\begin{align}
A^{\mu} & \to A^{\mu}+\partial^{\mu}\chi\,, \label{eq:fake_U(1)1}\\
\pi& \to \pi + m_{X}\chi\,, \label{eq:fake_U(1)2}
\end{align}
where $\chi(x)$ is the gauge parameter.
Under these transformations~\eqref{eq:fake_U(1)1} and \eqref{eq:fake_U(1)2}, $X^\mu$ in Eq.~\eqref{eq:X} is invariant.

Therefore, in a Lagrangian, a Proca mass contribution for $X^\mu$ is not incompatible with this $U(1)$ gauge symmetry. In fact, we can write
\begin{align}
    \mathcal{L} \supset \frac12\,m_X^2 X^\mu X_\mu=\frac{1}{2}(\partial_\mu\pi -m_X A_\mu)\,(\partial^\mu\pi-m_X A^\mu)\,.\label{eq:mass_pi}
\end{align}
This construction allows Lagrangian terms such as
\begin{align}
    \mathcal{L}\supset \frac{\bar{\lambda}_{HX}}{\Lambda^2}(\partial_\mu\pi -m_X A_\mu)\,(\partial^\mu\pi-m_X A^\mu)H^\dagger H\,.\label{eq:lambda_HX_6}
\end{align}
The mass term in Eq.~\eqref{eq:mass_pi} provides a kinetic term for $\pi$. Note that no kinetic term for $\pi$ comes instead from the term $-(1/4)\,{F^X}_{\mu\nu}{F^X}^{\mu\nu}$, since $F^X_{\mu\nu}=\partial_\mu X_\nu-\partial_\nu X_\mu=\partial_\mu A_\nu-\partial_\nu A_\mu$.
We can make use of the $U(1)$ gauge redundancy to fix $\pi=0$ (unitary gauge).
In this gauge, the gauge-invariant dimension-6 parent reduces to the nominal
dimension-4 interaction
\begin{align}
\mathcal{L}_{\rm X} = \lambda_{HX} (H^\dagger H) X_\mu X^\mu,
\end{align}
with the mass-suppressed coefficient relation
\begin{align} 
\lambda_{HX}=\bar\lambda_{HX} (m_X^2/\Lambda^2). 
\label{eq:bar_lambda}
\end{align}
Along the same lines, we can introduce a dimension-8 operator as follows:
\begin{align}
\mathcal{L}\supset{}&
2\frac{\bar{C}_{X}}{\Lambda^4}
(\partial_\mu\pi-m_XA_\mu)(\partial^\mu\pi-m_XA^\mu)
|D_\nu H|^2 .
\label{eq:C_X_bar}
\end{align}
We will consider this operator in more detail below. After fixing the
$\pi=0$ gauge, it becomes
$(C_X/\Lambda^2)|D_\mu H|^2X_\nu X^\nu$, with
$C_X=2\bar C_Xm_X^2/\Lambda^2$.

Note that once the SM Higgs acquires a vacuum expectation value $v = 246~{\rm GeV}$, the dark photon mass is shifted as
\begin{align}
m_X^2 \to m_X^2 + \lambda_{HX} v^2.
\end{align}
For freeze-in DM production near the reheating temperature, we assume that the electroweak symmetry is unbroken,
so that the DM mass originates solely from the Stueckelberg mass term during this epoch.
To avoid a negative mass squared in the late Universe, we require
\begin{align}
\lambda_{HX} \geq 0
\quad \text{or} \quad
\left( \lambda_{HX} < 0 \ \text{and} \ m_X^2 > |\lambda_{HX}| v^2 \right).
\label{eq:neg_lambda}
\end{align}

A notable feature of the dimension-4 Higgs-portal Stueckelberg operator is that 
the scattering amplitude for longitudinal modes grows with the center-of-mass energy 
as powers of $\sqrt{s}/m_X$.  
This growth leads to a violation of perturbative unitarity at high energies, which results in the condition~\cite{Kribs:2022gri}:
\begin{align}
s_{\rm max} \simeq \frac{m_X^2}{|\lambda_{HX}|} . \label{eq:per_uni}
\end{align}
Correspondingly, the longitudinal mode becomes strongly coupled at the scale
\begin{align}
\Lambda_{X} \simeq \frac{m_X}{\sqrt{|\lambda_{HX}|}}\,, \label{eq:UV_stu}
\end{align}
which must be taken into account when analyzing Higgs-portal Stueckelberg DM production.  
This behavior contrasts with the dimension-4 Higgs-portal scalar DM operator $(H^\dagger H)\phi^2$ involving a scalar DM field $\phi$, 
for which the scattering amplitude does not exhibit any growth with $s$.

We also consider the following dimension-6 Higgs-portal Stueckelberg operators, which contribute to the 2-to-2 scattering of a Higgs pair into a DM pair in the electroweak-symmetric phase, in addition to Eqs.~\eqref{eq:op1} and \eqref{eq:op2}:
\begin{align}
\mathcal{O}_{X} &= |D_\mu H|^2\, X_\nu X^\nu, \label{eq:op1_stu} \\
\tilde{\mathcal{O}}_{X} &= i(H^{\dagger} \overleftrightarrow{D^\mu} H)\, \tilde{X}_{\mu\nu} X^{\nu}. \label{eq:op2_stu}
\end{align}
Their Wilson coefficients are denoted by $C_X/\Lambda^2$ and
$\tilde C_X/\Lambda^2$, respectively.
The operators
\begin{equation}
(\partial^\mu\partial^\nu|H|^2)X_\mu X_\nu,\qquad
i(H^\dagger\overleftrightarrow{D^\mu}H)X_{\mu\nu}X^\nu,\qquad
(H^\dagger D_\mu H)X^\nu\partial^\mu X_\nu+{\rm h.c.}
\label{eq:redundant_stu}
\end{equation}
are redundant upon use of the field equations of motion.

Following the Stueckelberg formulation of dark photon EFT discussed in Ref.~\cite{Song:2023lxf}, one may also construct operators of the form
\begin{equation}
H^\dagger (D^\mu D^\nu H)\, X_\mu X_\nu + \mathrm{h.c.}
\label{eq:spin2_dim8}
\end{equation}
Equation~\eqref{eq:spin2_dim8} is an independent dimension-6 operator in unitary gauge.
We set its independent Wilson coefficient to zero as part of our EFT benchmark assumption.
As shown explicitly in Appendix~\ref{app:tildeOX}, the Levi-Civita tensor in
$\tilde{\mathcal O}_X$ eliminates the channel with two longitudinal vectors,
leaving final states with one longitudinal and one transverse vector. For equal unitary-gauge coefficients,
its squared amplitude is suppressed by $\mathcal O(m_X^2/s)$ relative to that of
$\mathcal O_X$. This comparison alone, however, does not justify omitting the
operator because the two coefficients have different natural Stueckelberg power
counting. We therefore retain $\tilde{\mathcal O}_X$ in the EFT and in the
general freeze-in yield below. It produces a positive $T_{\rm reh}^5$ contribution
that does not interfere with the diagonal $\phi_i\phi_i$ channels.

The full effective Lagrangian considered in this work is
\begin{align}
\mathcal{L}_\mathrm{VDM}
= \frac{m_X^2}{2}\, X_\mu X^\mu
 + \frac{C}{\Lambda^2}\, \mathcal{O}
 + \frac{\tilde{C}}{\Lambda^2}\, \tilde{\mathcal{O}} 
 + \lambda_{HX} (H^\dagger H) X_\mu X^\mu
 + \frac{C_X}{\Lambda^2}\, \mathcal{O}_X
 + \frac{\tilde C_X}{\Lambda^2}\,\tilde{\mathcal O}_X. \label{eq:lag2}
\end{align}

For the dimension-6 operator in~\eqref{eq:op1_stu}, we estimate the associated
longitudinal strong-coupling scale by requiring that the squared scattering
amplitude for Higgs-pair to DM-pair production become of order unity. This leads to
\begin{align}
\Lambda^{{\rm dim-6}}_X \simeq
\frac{2^{1/4}(\Lambda m_X)^{1/2}}{|C_X|^{1/4}}.
\label{eq:UV_stu2}
\end{align}
For $\tilde{\mathcal O}_X$, a gauge-invariant parent interaction is
\begin{align}
\mathcal L\supset
-\frac{\bar{\tilde C}_X}{\Lambda^3}
i(H^\dagger\overleftrightarrow{D^\mu}H)\tilde X_{\mu\nu}
(\partial^\nu\pi-m_XA^\nu).
\label{eq:tilde_parent_stu}
\end{align}
In the $\pi=0$ gauge this gives
$\tilde C_X=\bar{\tilde C}_Xm_X/\Lambda$, or equivalently
$\bar{\tilde C}_X=\tilde C_X\Lambda/m_X$. Its high-energy amplitude
scales as $\bar{\tilde C}_X s^{3/2}/\Lambda^3$, so its corresponding
strong-coupling scale is of order
\begin{align}
\Lambda_X^{\widetilde{\mathcal O}}\sim
\frac{\Lambda}{|\bar{\tilde C}_X|^{1/3}}
=\left(\frac{\Lambda^2m_X}{|\tilde C_X|}\right)^{1/3}.
\label{eq:UV_stu_tilde}
\end{align}

One possible weakly coupled ultraviolet (UV) completion of the Higgs-portal
Stueckelberg DM scenario introduces a dark Higgs boson. Its radial mode restores
perturbative unitarity in longitudinal vector-boson scattering at high energies
~\cite{Kribs:2022gri}. More generally, however, the scale $\Lambda$ suppressing the
field-strength operators and the strong-coupling scales $\Lambda_X$ and
$\Lambda_X^{{\rm dim-6}}$ (and the analogous scale
$\Lambda_X^{\widetilde{\mathcal O}}$ for $\tilde{\mathcal O}_X$) have distinct
physical origins. Perturbative unitarity
alone does not determine their ordering. The combined description is valid only
below the smallest of these scales,
$\Lambda_{\rm eff}\equiv
\min\{\Lambda,\Lambda_X,\Lambda_X^{{\rm dim-6}},
\Lambda_X^{\widetilde{\mathcal O}}\}$, and the freeze-in calculation
requires the characteristic production energy to be below $\Lambda_{\rm eff}$.

To organize all operators in Eq.~\eqref{eq:lag2} within a conservative
single-cutoff EFT, we adopt the additional power-counting assumption~\cite{Bertuzzo:2024bwy}
\begin{align}
\Lambda_{X},\ \Lambda^{{\rm dim-6}}_X,\ 
\Lambda_X^{\widetilde{\mathcal O}} > \Lambda. \label{eq:cutoff_cond}
\end{align}
Thus, Eq.~\eqref{eq:cutoff_cond} is an assumption about matching to a common UV
description, rather than a model-independent consequence of unitarity.
From Eq.~\eqref{eq:cutoff_cond} together with Eqs.~\eqref{eq:UV_stu} and \eqref{eq:UV_stu2}, respectively, 
we obtain the following conditions:
\begin{align}
|\lambda_{HX}| &< \frac{m^2_X}{\Lambda^2},\label{eq:coeff_cond}\quad{\rm or}\quad|\bar{\lambda}_{HX}|<1\,,\\
|C_X| &< 2\frac{m^2_X}{\Lambda^2}, \quad{\rm or}\quad|\bar{C}_{X}|<1,\label{eq:coeff_cond2}\\
|\tilde C_X|&<\frac{m_X}{\Lambda},\quad{\rm or}\quad
|\bar{\tilde C}_X|<1. \label{eq:coeff_cond_tilde}
\end{align}
Within this single-cutoff power counting, the barred coefficients are smaller
than order unity. Separately, we assume $m_X<\Lambda$ so that the vector is a
degree of freedom of the EFT.
Note that Eq.~\eqref{eq:neg_lambda} is automatically satisfied by the condition~\eqref{eq:coeff_cond}, since $\Lambda > v$.
If the hierarchy~\eqref{eq:cutoff_cond} is not imposed, the bounds
\eqref{eq:coeff_cond}--\eqref{eq:coeff_cond_tilde} do not follow. In that case,
$\Lambda_{\rm eff}$, rather than $\Lambda$, controls the range of validity, and
Stueckelberg-allowed interactions can dominate even for $C,\tilde C=\mathcal O(1)$,
provided that the production energies remain below $\Lambda_{\rm eff}$. The numerical
analysis below focuses on the common-cutoff benchmark~\eqref{eq:cutoff_cond}; a detailed
study of multiscale UV completions is left for future work.

%%%%%%%%%%%%%
%%%%%%%%%%%%%
\section{DM relic density via Higgs-Portal Freeze-in} \label{sec:dm_relic}
In this section, we examine the constraints on the Higgs-portal couplings relevant for the freeze-in production of DM.  
First, in Sec.~\ref{sec:dm_relic_setup}, we briefly review the method used to compute DM freeze-in production, following Ref.~\cite{Kim:2023bbs} 
by Kim, Lee, and one of the authors (K.~Y.).  
Next, in Sec.~\ref{sec:dm_relic_dim6}, we present the viable parameter space that reproduces the observed DM relic density using the dimension-6 operators in Eqs.~\eqref{eq:op1}--\eqref{eq:lag}.  
Then, in Sec.~\ref{sec:dm_relic_stu}, we introduce the dimension-4 and
dimension-6 operators permitted by the Stueckelberg construction and identify
the power-counting assumptions under which they are subdominant to the
field-strength portals.
Finally, we consider the case in which the Wilson coefficients of the
field-strength operators are suppressed below unity and discuss the DM relic
density including the Stueckelberg-allowed operators.
\subsection{Setup for Freeze-in DM}
\label{sec:dm_relic_setup}
In this subsection, we briefly review the method for computing the DM
freeze-in production following Refs.~\cite{Kim:2023bbs,Mambrini2021}.  
In our model, DM is produced through Higgs-portal interactions in the
electroweak-symmetric phase. All four real scalar components of the Higgs
doublet participate through diagonal and distinct-component channels.

The relevant Higgs-portal operators are given by
\begin{align}
&\frac{C}{\Lambda^{2}}(H^\dagger H)X_{\mu\nu}X^{\mu\nu},
\qquad
\frac{\tilde C}{\Lambda^{2}}(H^\dagger H)
\tilde X_{\mu\nu}X^{\mu\nu},
\qquad
\lambda_{HX}(H^\dagger H)X_\mu X^\mu,\nonumber\\
&\frac{C_X}{\Lambda^2}|D_\mu H|^2X_\nu X^\nu,
\qquad
\frac{\tilde C_X}{\Lambda^2}
i(H^\dagger\overleftrightarrow{D^\mu}H)\tilde X_{\mu\nu}X^\nu .
\end{align}
The first two terms correspond to the gauge-invariant dimension-6 CP-even and CP-odd Higgs-portal operators, 
whereas the last three terms are the dimension-4 and dimension-6 operators allowed by the
Stueckelberg mechanism.

\begin{figure}[!t]
\begin{center}
 \includegraphics[width=0.25\textwidth,clip]{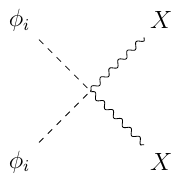}
\end{center}
\caption{Feynman diagrams for DM production from scalar-pair annihilation within the Higgs doublet.
Here, $\phi_i\,(i=1,2,3,4)$ denote 
the four real scalar components of the Higgs doublet, and $X$ represents the real vector DM.}
\label{fig:diagram_relic}
\end{figure}
DM is produced via scalar-scalar scattering within the Higgs doublet. The
field-strength, $\lambda_{HX}$, and $\mathcal O_X$ interactions contribute to
the diagonal process $\phi_i\phi_i\to XX$ shown in
Figure~\ref{fig:diagram_relic}, whereas $\tilde{\mathcal O}_X$ contributes to
the distinct-component channels $\phi_1\phi_2\to XX$ and
$\phi_3\phi_4\to XX$. Here,
$\phi_i\, (i=1,2,3,4)$ are the four real scalar components of the Higgs doublet,
\begin{align}
H = \frac{1}{\sqrt{2}}
\begin{pmatrix}
\phi_1 + i\phi_2 \\
\phi_3 + i\phi_4
\end{pmatrix}.
\end{align}

In the freeze-in mechanism~\cite{Dodelson:1993je,Hall:2009bx,Bernal:2017kxu}, the DM must remain out of thermal equilibrium.
This condition can be expressed as the reaction rate being smaller than the Hubble expansion rate~\cite{Bernal:2018kcw},
\begin{align}
n_\phi \langle \sigma v \rangle < H, \label{eq:freezein}
\end{align}
where $n_\phi$ is the number density of particles in the thermal bath, $\langle \sigma v \rangle$ is the thermally averaged cross section, and $H$ is the Hubble parameter.
For later quantitative checks, we define the particle-production rate per
equilibrium bath particle,
${\cal R}_{\rm FI}(T)\equiv2\gamma(T)/(n_\phi^{\rm eq}H)$,
where $n_\phi^{\rm eq}=g_\phi\zeta(3)T^3/\pi^2$ for the four real Higgs
components and $\gamma$ is the event density defined in
Eq.~\eqref{eq:rate}. For the UV-dominated interactions considered here, the
freeze-in condition ${\cal R}_{\rm FI}<1$ is most restrictive at
$T\simeq T_{\rm reh}$. In
Appendix~\ref{app:consistency}, we bound this quantity analytically on the
relic-density contours and find
${\cal R}_{\rm FI}(T)<3.4\times10^{-2}$ throughout
$m_h\leq T\leq T_{\rm reh}$ for the displayed $\Omega_Xh^2=0.12$ contours after
applying the operator-matched conservative structure-formation reference
scales of Appendix~\ref{app:lyman}.

For the Higgs-portal Stueckelberg DM, the scattering amplitude for $\phi_i \phi_i \to XX$ can be roughly estimated by considering the two contributions separately (neglecting interference terms):
\begin{align}
|{\cal M}_{\phi_i \phi_i \to XX}|^2 \sim \lambda^2_{HX} \frac{s^2}{m_X^4},\quad
|{\cal M}_{\phi_i \phi_i \to XX}|^2 \sim C_X^2\frac{s^4}{\Lambda^4 m_X^4} ,
\end{align}
where $s$ is the Mandelstam variable, which is roughly $s \sim T^2$ for a thermal bath at temperature $T$.
The corresponding thermally averaged cross sections are then estimated as
\begin{align}
\langle \sigma v \rangle \sim \frac{|{\cal M}|^2}{s} \sim
\frac{\lambda_{HX}^2 T^2}{m_X^4}, \quad
\langle \sigma v \rangle \sim \frac{|{\cal M}|^2}{s}\sim\frac{C_X^2 T^6}{\Lambda^4 m_X^4}.
\end{align}
Assuming a relativistic thermal bath, $n_\phi \sim T^3$, and comparing with the Hubble expansion rate $H \sim T^2/M_{\rm pl}$, the freeze-in condition~\eqref{eq:freezein}
leads to the following upper bounds on the couplings:
\begin{align}
|\lambda_{HX}| \lesssim \frac{m^2_X}{(T_{\rm reh}^{3} M_{\rm pl})^{1/2}},
\label{eq:stu-fimp0}\\
|C_X|/\Lambda^{2} \lesssim \frac{m^2_X}{(T_{\rm reh}^7M_{\rm pl})^{1/2}},
\label{eq:stu-fimp}
\end{align}
where $T_{\rm reh}$ is the reheating temperature and $M_{\rm pl} = 2.435\times 10^{18}~{\rm GeV}$ is the reduced Planck mass.
Here we take the reheating temperature as the scale for UV freeze-in in our scenario.
In this context, UV freeze-in refers to DM production that occurs predominantly at very high temperatures, close to the reheating scale, where the effective interactions are generated.
For the retained parity-odd Stueckelberg operator, whose high-energy squared
amplitude scales as
$|\tilde C_X|^2s^3/(\Lambda^4m_X^2)$, the analogous estimate is
\begin{align}
\frac{|\tilde C_X|}{\Lambda^2}
\lesssim\frac{m_X}{(T_{\rm reh}^5M_{\rm pl})^{1/2}}.
\label{eq:stu-fimp-tilde}
\end{align}

Similarly, the gauge-invariant dimension-6 Higgs-portal operators are constrained by~\cite{Kim:2023bbs}
\begin{align}
|C|/\Lambda^{2},\, |\tilde{C}|/\Lambda^{2} \lesssim (T_{\rm reh}^{3} M_{\rm pl})^{-1/2}. \label{eq:dim6-fimp}
\end{align}

For gauge-invariant dimension-6 operators with order-unity Wilson coefficients,
the out-of-equilibrium condition required for freeze-in gives the following order-of-magnitude lower bound:
\begin{align}
\Lambda \gtrsim \left( T_{\rm reh}^{3} M_{\rm pl} \right)^{1/4}. \label{eq:cutoff_fimp}
\end{align}
For example, this implies 
$\Lambda \gtrsim 10^{15}~{\rm GeV}$ for $T_{\rm reh} = 10^{14}~{\rm GeV}$ 
and 
$\Lambda \gtrsim 10^{9}~{\rm GeV}$ for $T_{\rm reh} = 10^{6}~{\rm GeV}$.  
For the illustrative temperatures given above, Eq.~\eqref{eq:cutoff_fimp} places the minimum ratio $\Lambda/T_{\rm reh}$ at roughly $10$--$10^3$, assuming order-unity field-strength coefficients.  
In this regime, the field-strength EFT-validity condition
$T_{\rm reh}\lesssim\Lambda$ is automatically satisfied in the limit where
the DM and Higgs masses can be neglected.
Equations~\eqref{eq:stu-fimp0} and \eqref{eq:stu-fimp} are automatically satisfied by Eqs.~\eqref{eq:coeff_cond} and \eqref{eq:coeff_cond2}, respectively, together with Eq.~\eqref{eq:cutoff_fimp}.
The parity-odd estimate above is likewise satisfied by
Eq.~\eqref{eq:coeff_cond_tilde} together with Eq.~\eqref{eq:cutoff_fimp}.

For the dimension-4 and dimension-6 Stueckelberg operators, the corresponding
strong-coupling scales are automatically higher than the reheating temperature
under the freeze-in out-of-equilibrium conditions:
\begin{align}
\Lambda_X \gtrsim (T_{\rm reh}^3 M_{\rm pl})^{1/4} > T_{\rm reh}.\\
\Lambda^{{\rm dim-6}}_X \gtrsim 2^{1/4}
(T_{\rm reh}^7 M_{\rm pl})^{1/8} > T_{\rm reh}.
\\
\Lambda_X^{\widetilde{\mathcal O}}
\gtrsim(T_{\rm reh}^5M_{\rm pl})^{1/6}>T_{\rm reh}.
\label{eq:cutoff_fimp_tilde}
\end{align}
These inequalities by themselves only show that the strong-coupling scales
exceed $T_{\rm reh}$; they do not guarantee control over the high-energy tail
of the thermal integral. In the numerical analysis we therefore impose the
stronger characteristic-energy criterion described in
Sec.~\ref{sec:cosmology}.
Under the common-cutoff assumption~\eqref{eq:cutoff_cond}, we additionally
impose Eqs.~\eqref{eq:coeff_cond}--\eqref{eq:coeff_cond_tilde}. Without that
assumption, these coefficient bounds do not follow; the necessary condition is
instead that the characteristic production energy remain below
$\Lambda_{\rm eff}$.

We consider a homogeneous 
and isotropic cosmological 
background described by the line element 
${\rm d}s^2 = {\rm d}t^2 - a^2(t) \, 
\delta_{ij} \, {\rm d}x^i {\rm d}x^j$, 
where $a(t)$ is the scale factor that depends 
on the cosmic time $t$.
The number density of DM obeys the Boltzmann equation,
\begin{align}
\dot n_X + 3H n_X = 2\gamma(T),
\label{eq:boltzmann}
\end{align}
where  a dot denotes the derivative with respect to $t$, 
$\gamma(T)$ is the production-event density per unit volume and per unit time, 
which depends on the background temperature $T$,
and $H \equiv \dot{a}/a$ is the Hubble expansion rate.
The DM production occurs during radiation dominance, in which $H$ 
can be expressed as
\begin{align}
H=\sqrt{\frac{\rho_R}
{3M_{\rm pl}^2}}
=\sqrt{\frac{g_*}{90}}
\frac{\pi T^2}{M_{\mathrm{pl}}}\,,
\label{Hra}
\end{align}
where $\rho_R=\pi^2 g_* T^4/30$ is 
the radiation energy density, 
and $g_*$ is the number 
of effective relativistic degrees 
of freedom (DOFs). 
Introducing the DM yield $Y_X\equiv n_X/T^3$ and using 
the property $T \propto a^{-1}$, we obtain
\begin{align}
\frac{{\rm d}Y_X}{{\rm d}T}
   = -\,\frac{2\gamma(T)}{H(T)\,T^{4}} .
\label{eq:boltzmann_yield}
\end{align}
For a general scattering process $A B \to C D$, the production-event density is given by
\begin{align}
\gamma(T) = g_\phi\int f_A f_B \frac{E_A E_B \mathrm{d}E_A \mathrm{d}E_B \mathrm{d}\cos{\theta_{AB}}}{1024\pi^6}\int|\mathcal{M}_{AB\to CD}|^2
\mathrm{d}\Omega_{AC},
\label{eq:rate}
\end{align}
where 
\begin{align}
f_{A, B}  = \frac{1}{e^{E_{A, B}/T} -1}
\label{eq:bose_d}
\end{align} 
are the Bose--Einstein distributions for the Higgs bosons with the corresponding 
energies $E_A$ and $E_B$, respectively.
Here, $g_\phi = 4$ is the number of Higgs DOFs,
$\theta_{AB}$ is the angle between the incoming Higgs bosons $A$ and $B$ in the laboratory frame,
$\mathrm{d}\Omega_{AC}$ is the solid angle of one of the outgoing particles in the center-of-mass (COM) frame,
and $|{\cal M}_{A B \to C D}|^2$ denotes the squared matrix element of the process $AB \to CD$.

The Higgs field is kept in thermal equilibrium by its SM gauge and Yukawa
interactions. In the electroweak-symmetric plasma it also acquires a thermal
mass $m_{H,{\rm th}}^2=c_H T^2$, where
$c_H=(3g_2^2+g_1^2)/16+y_t^2/4+\lambda_H/2$ at leading order
~\cite{Lebedev:2019ton}. We neglect this mass in the analytic rates.
Appendix~\ref{app:consistency} gives a quantitative check. Taking the
conservative value $c_H=0.40$ and the same Bose--Einstein statistics as in
Eq.~\eqref{eq:bose_d}, the rate changes by no more than about six percent for
all retained $F_3$, $F_5$, and $F_7$ contributions. It therefore does not
alter the powers of $T_{\rm reh}$ or the hierarchy among the contributions.

In Eq.~\eqref{eq:rate}, we neglect the overall 2-body phase-space factor 
$\sqrt{1 - 4 m_X^2/s}$. 
This is justified because, as we show in the next sections, the DM yield is dominantly produced around 
the reheating temperature for the parameter space of interest and when $m_X \ll T_{\rm reh}$, 
the overall 2-body phase-space factor can be approximated as unity.
If the DM mass is larger than the temperature,
$E_A, E_B > m_X > T$, the DM production is exponentially suppressed due to the Boltzmann factors 
$f_{A,B} \simeq \exp(-E_{A,B}/T)$.
Apart from the sub-six-percent thermal-mass correction quantified above, we
treat the initial-state Higgs excitations as massless in the
electroweak-symmetric phase.

The Mandelstam variables, $s$ and $t$, are related to the angle between the initial states in the laboratory frame, $\theta_{AB}$, and the scattering angle $\theta_{AC}$ in the COM frame, as follows:
\begin{align}
s &= 2 E_A E_B \left( 1 - \cos\theta_{AB} \right), \label{ss} \\
t &= \frac{s}{2} \left( \sqrt{1 - \frac{4 m_X^2}{s}} \cos\theta_{AC} - 1 \right) + m_X^2.
\end{align}
We can take $\mathrm{d}\Omega_{AC} = 2 \pi \, \mathrm{d}\cos\theta_{AC}$ due to the azimuthal symmetry in the COM frame.
Here, we note that $s$ appearing in $t$ is taken in the COM frame, but its value is identical to the one in the laboratory frame.
Thus, we use the form of $s$ in the laboratory frame in Eq.~\eqref{ss}, which is appropriate for thermal averages.

The relic density is obtained by integrating Eq.~\eqref{eq:boltzmann_yield} 
from the reheating temperature down to temperatures of order the electroweak
crossover. The broken-phase contribution below that scale is negligible for
the UV-dominated parameter space considered here.
\subsection{Gauge-Invariant Dimension-6 Operators} \label{sec:dm_relic_dim6}

First, in terms of the coefficients of the gauge-invariant Higgs-portal dimension-6 operators, 
we obtain the squared scattering amplitude for 
$\phi_i \phi_i \to XX$ ($i = 1,2,3,4$) by implementing the model in 
FeynRules~\cite{Christensen:2008py,Alloul:2013bka,Christensen:2009jx}
and using CalcHEP~\cite{Belyaev:2012qa}. 
The result is given by
\begin{align}
|\mathcal{M}_{\phi_i \phi_i \to XX}|^2
= \frac{2}{\Lambda ^4} 
\left[C^2(s^2 -4 m^2_X s + 6 m^4_X)+ \tilde{C}^2 s (s - 4 m^2_X)\right],
\label{eq:amp1}
\end{align}
where the symmetry factors for identical particles in both the initial state
($\phi_i \phi_i$) and the final state ($XX$) are included.
The polarization states of the final-state DM ($XX$) are summed over.
Both operators yield terms proportional to $s^2$ in
Eq.~\eqref{eq:amp1}.
In the high-energy limit, where $s \gg m_X^2$,  the squared scattering amplitude in Eq.~\eqref{eq:amp1} can be approximated as
\begin{align}
|\mathcal{M}_{\phi_i \phi_i \to XX}|^2
\simeq \frac{2s^2}{\Lambda^4}(C^2 + \tilde{C}^2).
\label{eq:amp1_highE}
\end{align}
Compared to Appendix~A of Ref.~\cite{Kim:2023bbs}, 
we find the correspondences $c_{2,0} = 2 (C^2 + \tilde{C}^2)$ and $M = \Lambda$. 
Therefore, the production-event density can be obtained from Eq.~\eqref{eq:amp1_highE} as
\begin{align}
\gamma(T) &= g_\phi \frac{c_{2,0}}{M^4} \frac{\pi^3 T^8}{5400} \\
&= g_\phi\frac{C^2 + \tilde{C}^2}{\Lambda^4} \frac{\pi^3 T^8}{2700}.
\label{eq:r1}
\end{align}
From Eqs.~(\ref{Hra}) 
and (\ref{eq:r1}), 
the right-hand side of Eq.~(\ref{eq:boltzmann}) 
can be expressed as a function of $T$. 
We are interested in DM production from 
the end of reheating, characterized by 
the temperature 
$T_{\rm reh}$, 
down to a reference temperature of order the electroweak crossover,
$T\simeq m_h$. Below that scale the Higgs d.o.f.\ and the
appropriate broken-phase rates change. Because the present production is UV
dominated, this low-temperature contribution is suppressed at least by
$(m_h/T_{\rm reh})^3$ and is negligible for the benchmarks studied here.
We therefore denote the asymptotic yield by $Y_X(m_h)$.
Since the integration of the Boltzmann equation is dominated by high temperatures for the dimension-6 operators considered here, we take 
$g_*$ to be its value at $T_{\rm reh}$, i.e., $g_* = g_*(T_{\rm reh})$, which applies to the UV freeze-in regime with $T_{\rm reh} \gg m_h$.
We use $g_*(T_{\rm reh})= 106.75$ for the SM. 
This justifies working in the electroweak symmetric phase.

Integrating Eq.~(\ref{eq:boltzmann_yield}) with respect to $T$,
we obtain 
\begin{align}
Y_X(T)=-2\times\frac{g_\phi}{\sqrt{g_*(T_{\rm reh})}}
\frac{\pi^2M_{\rm pl}(C^2 + \tilde{C}^2)}{270\sqrt{10}\,\Lambda^4}T^3+C_{\rm int}\,,
\end{align}
where $C_{\rm int}$ is an integration constant. 
The factor of 2 originates from the two DM particles produced in each event,
as displayed explicitly in Eq.~\eqref{eq:boltzmann}.
As the temperature $T$ decreases from 
$T_{\rm reh}$ to $m_h$, $Y_X$ 
increases. 
The constant $C_{\rm int}$ is determined by
the initial condition $Y_X(T)=Y_X(T_{\rm reh})$ 
at the end of reheating ($T=T_{\rm reh}$).
At the temperature $T=m_{h}$, the DM yield 
is then given by \cite{Kim:2023bbs}
\begin{align}
Y_X(m_h) &= Y_X(T_\mathrm{reh}) + \frac{g_\phi}{\sqrt{g_*(T_{\rm reh})}}\frac{\pi^2M_{\rm pl}(C^2 + \tilde{C}^2)}{135\sqrt{10}\Lambda^4}
(T^3_\mathrm{reh} - m^3_h).
\label{eq:yield1}
\end{align}
In the limit $T_\mathrm{reh} \gg m_h$ and assuming $Y_X(T_\mathrm{reh}) = 0$, Eq.~\eqref{eq:yield1} can be approximated as
\begin{align}
Y_X(m_h) \simeq \frac{g_\phi}{\sqrt{g_*(T_{\rm reh})}}\frac{\pi^2M_{\rm pl}(C^2 + \tilde{C}^2)}{135\sqrt{10}\Lambda^4}T^3_\mathrm{reh}.
\label{eq:yield1_fin}
\end{align}
The corresponding number density at the end of production is
$n_X(m_h)=Y_X(m_h)m_h^3$.  Once production has ceased, the comoving number
$n_Xa^3$ is conserved irrespective of whether $X$ is still relativistic at
$T=m_h$.  Its present-day rest-mass density is therefore
\begin{align}
\rho_{X}^{(0)}
=m_XY_X(m_h)m_h^3
\left[ \frac{a_0}{a(m_h)} \right]^{-3}\,,
\label{rhoX}
\end{align}
where $a_0$ and $a(m_h)$ are the scale factors 
today and at $T = m_h$, respectively.
Due to entropy conservation, 
$g_{*s} T^3 a^3 = \text{constant}$, where $g_{*s}$ denotes the effective entropy DOFs,
a comparison between $T = m_h$ and present temperature 
$T_0 \simeq 2.35 \times 10^{-13}\,\text{GeV}$ 
 (measured in the observations of cosmic microwave background) yields
\begin{align}
\left[ \frac{a_0}{a(m_h)} \right]^{-3}
\simeq \frac{g_{*s}(T_0)}
{g_{*s}(T_{\rm reh})} \frac{T_0^3}{m_h^3}\,,
\label{a0h}
\end{align}
where $g_{*s}(T_0) \simeq 3.91$ is the present-day 
effective entropy DOFs. Here we have used the approximation 
$g_{*s}(m_h) \simeq g_{*s} (T_{\rm reh})=106.75$
for the value of $g_{*s}$ at $T = m_h$. 

The present-day DM density parameter is 
defined as $\Omega_X = \rho_X^{(0)}/
(3 M_{\rm pl}^2 H_0^2)$, 
where $H_0$ is the present-day Hubble parameter, 
given by $H_0 = 2.13 \times 10^{-42}\, h$~GeV 
(with $h \approx 0.7$).
By using Eq.~(\ref{rhoX}) with Eqs.~(\ref{eq:yield1_fin}) and 
(\ref{a0h}), it follows that 
\begin{align}
\Omega_X h^2 &= 1.6\times 10^8 \left(\frac{m_X}{1~\mathrm{GeV}}\right)\left(\frac{g_{*s}(T_0)}{g_{*s}(T_{\rm reh})}\right) Y_X(m_h)\notag \\
&\simeq 2.3\times 10^6 \left(\frac{m_X}{1~\mathrm{GeV}}\right)\frac{\pi^2M_{\rm pl}(C^2 + \tilde{C}^2)}{135\sqrt{10}\Lambda^4}T^3_\mathrm{reh}\notag\\
&\simeq 0.12 \left(\frac{m_X}{100~\mathrm{GeV}}\right)\left(\frac{T_\mathrm{reh}}{10^{10}~\mathrm{GeV}}\right)^3
\left(\frac{10^{14}~\mathrm{GeV}}{\Lambda}\right)^4\times 1.1(C^2+\tilde{C}^2).
\label{eq:DM_relic1}
\end{align}
%Here, the effective entropy degrees of freedom are $g_{*s}(T_0) = 3.91$ at present and $g_{*s}(T_{\rm reh}) = 106.75$ at the reheating temperature.

\begin{figure}[!t]
\begin{center}
\includegraphics[width=0.40\textwidth,clip]{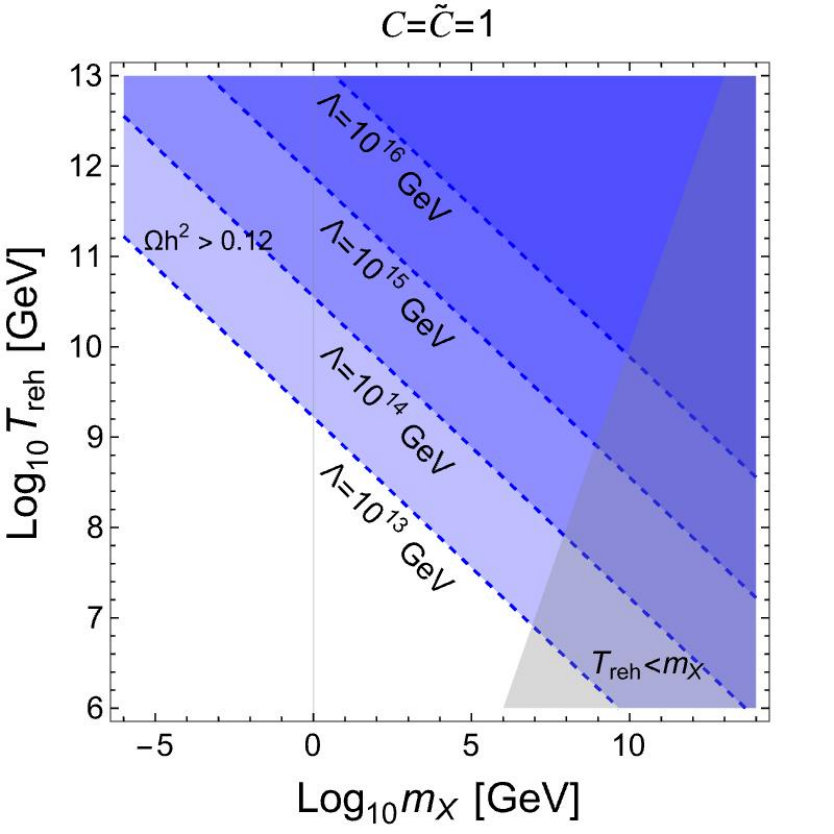}\,\,
\includegraphics[width=0.40\textwidth,clip]{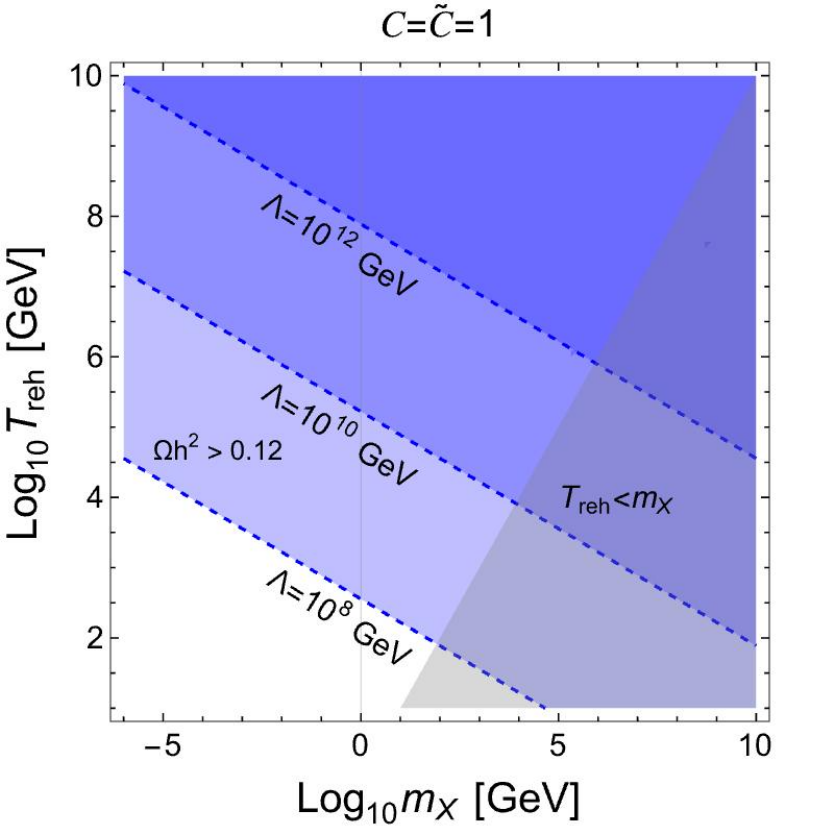}
\end{center}
\caption{
Parameter space in the $m_X$-$T_{\mathrm{reh}}$ 
plane consistent with the 
observed DM relic density for 
$C = \tilde{C} = 1$. 
In the left and right panels, 
we take the cutoff scale to be 
$\Lambda = 10^{13,14,15,16}~\mathrm{GeV}$ and 
$\Lambda = 10^{8,10,12}~\mathrm{GeV}$, respectively.  
The gray regions indicate $T_{\mathrm{reh}} < m_X$, 
where DM production is Boltzmann suppressed.  
The axes extend below the indicative structure-formation scale only to display
the analytic relic-density dependence; these portions are not claimed to be
cosmologically viable.
The DM relic density is overproduced in the blue region 
($\Omega_X h^2 > 0.12$), whereas the observed value 
is realized along the boundary.
}
\label{fig:RelicD}
\end{figure}
In Figure~\ref{fig:RelicD}, we show the parameter space in the $m_X$-$T_{\rm reh}$ plane that reproduces the observed DM relic density, 
$\Omega_X h^2 = 0.12$~\cite{Planck:2018vyg}.
For the Higgs-portal suppression scale, we take  
$\Lambda = 10^{13,14,15,16}~\mathrm{GeV}$ 
in the left panel and  
$\Lambda = 10^{8,10,12}~\mathrm{GeV}$ in the right panel,  
while keeping $C = \tilde{C} = 1$ fixed throughout.
In the gray regions, the condition for thermal scattering 
($m_X < T_{\mathrm{reh}}$) is not satisfied, resulting in Boltzmann-suppressed DM production.
The momentum distribution generated by freeze-in is nonthermal, so a thermal
warm-DM mass limit cannot be applied directly. For the benchmark
pure-freeze-in distribution studied in Ref.~\cite{Decant:2021mhj} at
$g_{*s}=106.75$, the inferred lower scales are a few keV with a conservative
Lyman-$\alpha$ calibration and about $15$~keV with a more stringent one. The
present UV $2\to2$ processes weight higher momenta more strongly. In
Appendix~\ref{app:lyman}, an rms-velocity mapping gives indicative
lower masses of $6$--$9$~keV for the conservative calibration and
$23$--$35$~keV for the stringent calibration, depending on whether the
$s^2$, $s^3$, or $s^4$ contribution dominates. These are reference scales,
not sharp exclusions. A definitive bound requires solving the
momentum-dependent Boltzmann equation and computing the corresponding matter
transfer function.
A larger cutoff scale $\Lambda$ or a lower reheating temperature $T_{\rm reh}$ generally favors a heavier 
DM mass $m_X$ to achieve $\Omega_X h^2 = 0.12$.
Moreover, a smaller cutoff scale $\Lambda$ 
requires a lower reheating temperature $T_{\rm reh}$  
to reproduce the correct DM abundance. 
By treating the reheating temperature as a free parameter, 
we find that a wide range of DM masses and cutoff scales 
is consistent with the observed DM relic density.
\begin{figure}[!t]
\begin{center}
\includegraphics[width=0.40\textwidth,clip]{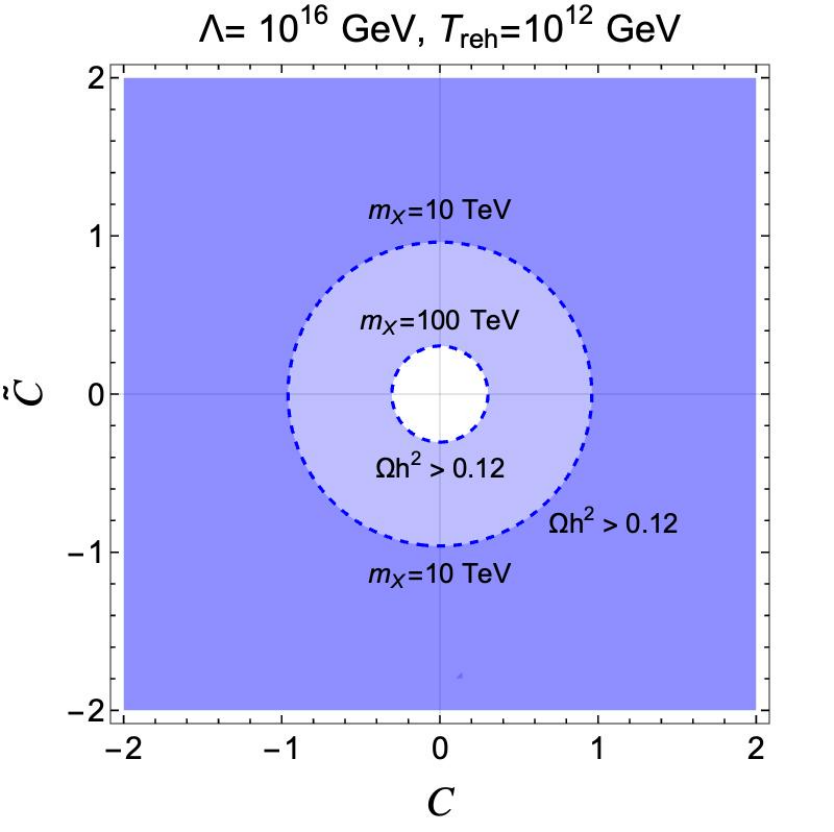} \,\,
\includegraphics[width=0.40\textwidth,clip]{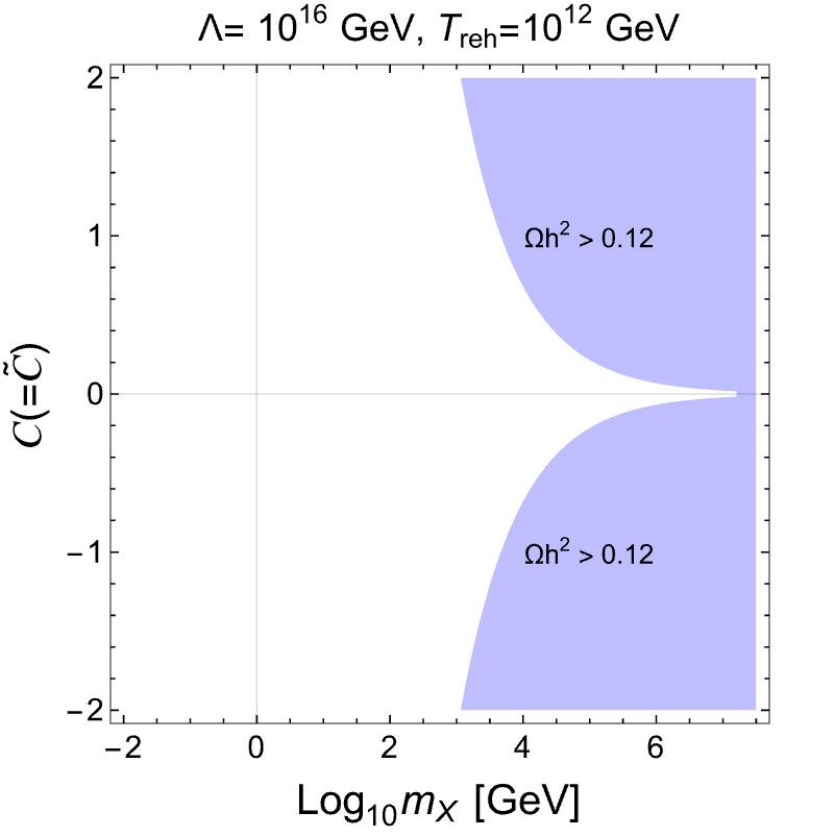}
\end{center}
\caption{Left: Parameter space in the $C$-$\tilde{C}$ plane.
We take $m_X = 10~\mathrm{TeV}$ and $100~\mathrm{TeV}$.  
Right: Parameter space 
in the $m_X$-$C~(=\tilde{C})$ plane.  
In all plots, we fix $\Lambda = 10^{16}~\mathrm{GeV}$ and $T_{\mathrm{reh}} = 10^{12}~\mathrm{GeV}$.}
\label{fig:RelicD2}
\end{figure}

In Figure~\ref{fig:RelicD2}, we present the parameter space in the 
$C$-$\tilde{C}$ plane (left) and in the $m_X$-$C\, (=\tilde{C})$ plane (right) 
that reproduces the observed DM relic density.  
Here, we fix $\Lambda = 10^{16}$~GeV and $T_{\mathrm{reh}} = 10^{12}$~GeV. 
In the regions outside the allowed boundaries, $\Omega_X h^2$ exceeds 0.12, 
while the observed value $0.12$ is realized along the boundaries.
In the left panel, for $m_X = 100$~TeV and $m_X = 10$~TeV, 
the boundaries corresponding to $\Omega_X h^2 = 0.12$ are given by the circles 
$C^2 + \tilde{C}^2 \simeq 0.3^2$ and $C^2 + \tilde{C}^2 \simeq 1^2$, respectively.  
The boundary curves in the right panel of Figure~\ref{fig:RelicD2} satisfy
$|C| \sqrt{m_X / (10^2~\mathrm{GeV})} \simeq 6.7$,
implying that $|C|$ decreases as $m_X$ increases.
\begin{figure}[!t]
\begin{center}
\includegraphics[width=0.40\textwidth,clip]{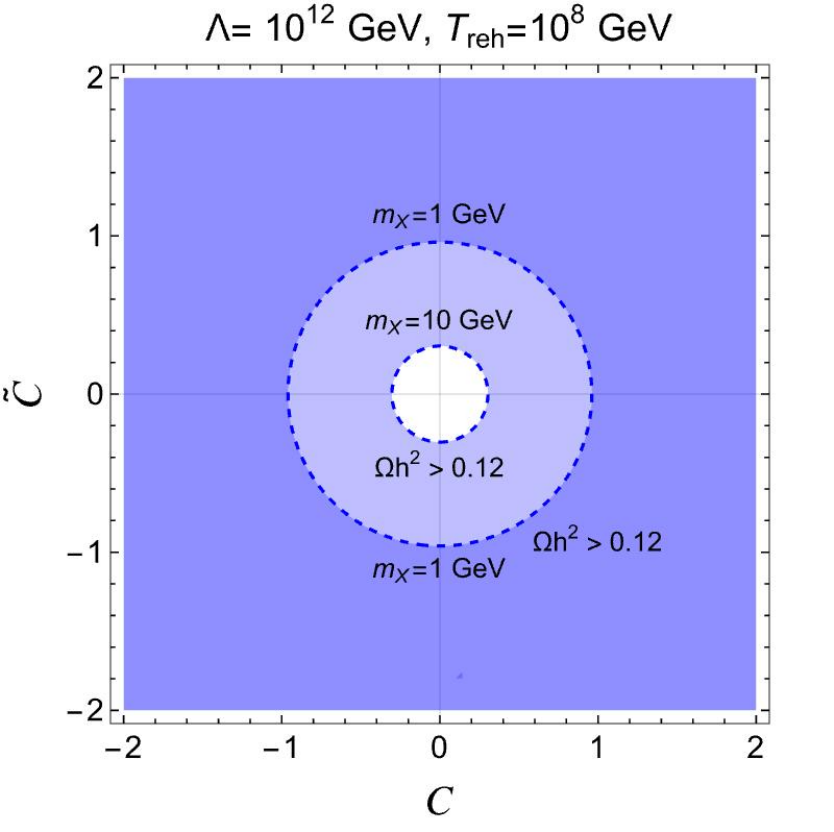} \,\,
\includegraphics[width=0.40\textwidth,clip]{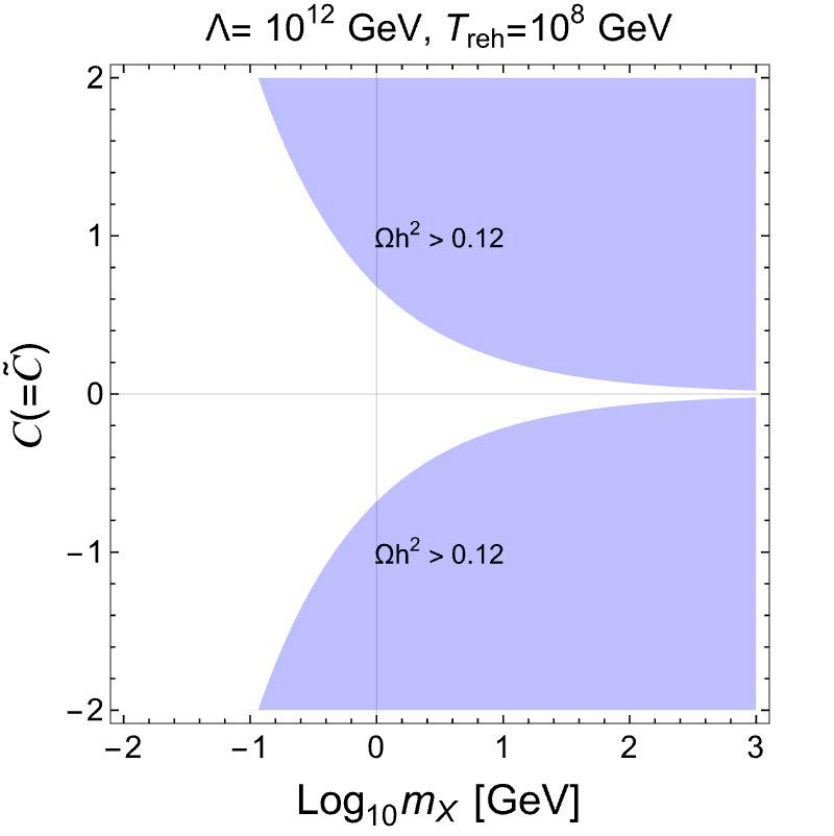}
\end{center}
\caption{Same as Figure~\ref{fig:RelicD2}, but with $\Lambda = 10^{12}$~GeV and $T_{\mathrm{reh}} = 10^{8}$~GeV.
We choose $m_X = 1~\mathrm{GeV}$ and $10~\mathrm{GeV}$ in the left panel.}
\label{fig:RelicD3}
\end{figure}
Figure~\ref{fig:RelicD3} shows the corresponding parameter space for 
$\Lambda = 10^{12}~\mathrm{GeV}$ and $T_{\mathrm{reh}} = 10^{8}~\mathrm{GeV}$.
In this case, the correct DM relic density is obtained around 
$m_X \simeq 1~\mathrm{GeV}$ for $(C^2+\tilde{C}^2)^{1/2} \simeq 1$, in agreement with the result 
in~Eq.~\eqref{eq:DM_relic1}.
Formally, the abundance relation extends into the keV range for lower $\Lambda$
and $T_{\rm reh}$, but only the part above the operator-dependent
Lyman-$\alpha$ limit can be regarded as viable.
\subsection{Stueckelberg Higgs-Portal Operators}
\label{sec:dm_relic_stu}

With the dimension-4 and dimension-6 Stueckelberg DM operators in addition to dimension-6 operators,
we obtain the squared scattering amplitude for 
$\phi_i \phi_i \to XX$ ($i = 1,2,3,4$) by implementing the model in 
FeynRules~\cite{Christensen:2008py,Alloul:2013bka,Christensen:2009jx}
and using CalcHEP~\cite{Belyaev:2012qa}:
\begin{align}
|{\cal M}_{\phi_i\phi_i\to XX}|^2&=\frac{2}{\Lambda^4}\left[C^2 \left(s^2-4 m_X^2 s+6 m_X^4\right)+\tilde{C}^2 s \left(s-4 m_X^2\right)\right]\nonumber\\
&\, \, \, \, +\left(\lambda_{HX}^2 +\frac{s^2}{4\Lambda ^4}C_X^2 -\lambda_{HX} C_X \frac{s}{\Lambda^2} \right)
\left[\frac{s(s-4 m_X^2)}{4m^4_X} + 3\right]\nonumber\\
&\, \, \, \, +\frac{3}{\Lambda^2} C\left(-2\lambda_{HX}+C_X\frac{s}{\Lambda^2}\right) \left(s-2 m_X^2\right),
\label{eq:amp_stu}
\end{align}
where the symmetry factors for identical particles in both the initial state
($\phi_i \phi_i$) and the final state ($XX$) are included.
The spin states of the final-state DM ($XX$) are summed over.
The Stueckelberg terms yield $s^2$ or higher powers in
Eq.~\eqref{eq:amp_stu}.
The operator $\tilde{\mathcal O}_X$ instead connects the distinct pairs
$\phi_1\phi_2$ and $\phi_3\phi_4$ to $XX$. Its amplitude, polarization
content, and reaction density are derived in Appendix~\ref{app:tildeOX}.
The energy growth of the amplitude from the dimension-4 Stueckelberg DM operator 
is consistent with Eq.~\eqref{eq:per_uni}.
From Eq.~\eqref{eq:amp_stu}, we observe an $s^2$ (or stronger) energy dependence, which is the same as that in Eq.~\eqref{eq:amp1} and implies UV freeze-in production.

In the high-energy limit, $s \gg m_X^2$, the squared scattering amplitude in Eq.~\eqref{eq:amp_stu} can be approximated as
\begin{align}
|\mathcal{M}_{\phi_i \phi_i \to XX}|^2
&\simeq \frac{2 s^2}{\Lambda^4} \left(C^2 + \tilde{C}^2\right)
+ \frac{s^2}{4 \Lambda^4} \left(\bar{\lambda}_{HX} - \bar{C}_X \frac{s}{\Lambda^2}\right)^2,
\label{eq:amp_stu_highE}
\end{align}
where the barred parameters are related to the original ones as
\begin{align} 
\bar{\lambda}_{HX} &= {\lambda}_{HX} \frac{\Lambda^2}{m_X^2}, \\
\bar{C}_X &= C_X \frac{\Lambda^2}{2m_X^2},\\
\bar{\tilde C}_X&=\tilde C_X\frac{\Lambda}{m_X}.
\end{align}
From Eqs.~\eqref{eq:coeff_cond}--\eqref{eq:coeff_cond_tilde}, these parameters satisfy
\begin{align}
|\bar{\lambda}_{HX}| < 1, \quad 
|\bar{C}_X| < 1,\quad
|\bar{\tilde C}_X|<1.
\label{eq:coeff_cond3}
\end{align}
These inequalities implement the common-cutoff power counting
\eqref{eq:cutoff_cond}; they are not required in a general multiscale EFT.

%One may wonder whether the term proportional to
%$\sim C C_X s^2/\Lambda^4 = C \bar{C}_X s^2 m_X^2/\Lambda^6$
%in Eq.~\eqref{eq:amp_stu} can be neglected in the high-energy limit.
%Indeed, it is subleading compared with the dominant contributions,
%$\sim C^2 s^2/\Lambda^4$ or $\sim \bar{C}_X^{\,2} s^4/\Lambda^8$,
%shown in Eq.~\eqref{eq:amp_stu_highE}.
%This can be easily checked in simple cases such as $C=0$, $\bar{C}_X=0$,
%$|C|=|\bar{C}_X|$, or $|C| > |C_X|$.
%The case with $|C_X| > |C|$ may be more subtle; however, it can be confirmed
%by parameterizing $|C| = \epsilon |C_X|$ with $\epsilon < 1$, and considering
%two regimes, $\epsilon \gg m_X^2/\Lambda^2$ and
%$\epsilon < f\, m_X^2/\Lambda^2$, where, for example, $f = 100$.
%Note that if $\epsilon \ge f m_X^2/\Lambda^2$, this automatically implies
%$\epsilon \gg m_X^2/\Lambda^2$.
%
In Eq.~\eqref{eq:amp_stu_highE},  we have ignored the mixing term of order 
 $C \bar{C}_X s^2 m_X^2/\Lambda^6$. This can be justified as follows.             
Taking the ratio of this term to the two terms 
of orders $C^2 s^2/\Lambda^4$ 
(or $\tilde{C}^2 s^2/\Lambda^4$) and 
$\bar{C}_X^2 s^4/\Lambda^8$ 
in Eq.~(\ref{eq:amp_stu_highE}), 
respectively, we obtain
\begin{equation}
r_1 \simeq \frac{\bar{C}_X}{C} 
\frac{m_X^2}{\Lambda^2}, \quad
r_2 \simeq \frac{C}{\bar{C}_X} 
\frac{\Lambda^2 m_X^2}{s^2}
\simeq \frac{1}{r_1} \frac{m_X^4}{s^2}\,.
\end{equation}
For nonzero $C$ and $\bar C_X$, the product of the two ratios is independent
of the Wilson-coefficient hierarchy:
\begin{equation}
|r_1r_2|\sim\frac{m_X^4}{s^2}\ll1
\qquad(s\gg m_X^2).
\label{CXcon}
\end{equation}
It follows immediately that
\begin{equation}
\min(|r_1|,|r_2|)
\leq\sqrt{|r_1r_2|}
\sim\frac{m_X^2}{s}\ll1.
\end{equation}
Thus, the mixed term is always subleading to at least one of the two positive
diagonal contributions. If either $C=0$ or $\bar C_X=0$, the mixed term
vanishes identically.

Within the common-cutoff benchmark, the gauge-invariant dimension-6 operators
dominate DM production when $C$ and $\tilde C$ are of order unity. Indeed,
the leading field-strength contribution in Eq.~\eqref{eq:amp_stu_highE} is
$2(C^2+\tilde C^2)s^2/\Lambda^4$, while the Stueckelberg terms are bounded by
the common-cutoff power counting and, for the higher-derivative terms, by
additional powers of $s/\Lambda^2<1$.

Next, we consider a different situation in which the coefficients $C$ and $\tilde{C}$ are smaller than order unity.
Such a hierarchy can arise from loop factors and symmetry-breaking spurions.
As an explicit matching pattern, consider a heavy real mediator $S$ of mass
$M_*\simeq\Lambda$ with derivative interactions
\begin{equation}
{\cal L}_{\rm int}=
\frac{g_H}{M_*}S|D_\mu H|^2+
\frac{g_X}{M_*}S(\partial_\mu\pi-m_XA_\mu)
(\partial^\mu\pi-m_XA^\mu).
\label{eq:mediator_matching}
\end{equation}
Integrating out $S$ at tree level generates the dimension-8 parent of
$\mathcal O_X$ in Eq.~\eqref{eq:C_X_bar}, with
$\bar C_X\sim g_Hg_X$. If an approximate derivative symmetry or mediator
selection rule forbids the nonderivative Higgs portal, the field-strength
operators, and the parity-odd Stueckelberg parent at tree level, then
$\bar\lambda_{HX}$, $C$, $\tilde C$, and $\bar{\tilde C}_X$ arise only
through a symmetry-breaking spurion $\epsilon$ and/or loops, giving
schematically
$\{\bar\lambda_{HX},C,\tilde C,\bar{\tilde C}_X\}/\bar C_X
\sim\epsilon/(16\pi^2)$. This provides a concrete mechanism for
$\bar C_X\gg
\{\bar\lambda_{HX},C,\tilde C,\bar{\tilde C}_X\}$. It is an illustrative
matching pattern, not a generic prediction of the Stueckelberg theory; a
complete UV completion is beyond the scope of this work.
Expanding Eq.~\eqref{eq:amp_stu_highE} and ordering the terms by powers of $s$, we obtain
\begin{align}
|\mathcal{M}_{\phi_i \phi_i \to XX}|^2
&\simeq \frac{s^4}{4 \Lambda^8} \bar{C}_X^2
- \frac{s^3}{2 \Lambda^6} \bar{\lambda}_{HX} \bar{C}_X
+ \frac{s^2}{\Lambda^4} \left[ 2 (C^2 + \tilde{C}^2) + \frac{1}{4} \bar{\lambda}_{HX}^2 \right].
\label{eq:amp_stu_highE_expanded2}
\end{align}
Compared to Appendix~A of Ref.~\cite{Kim:2023bbs}, the correspondences are
\begin{equation}
c_{4,0} = \frac{1}{4} \bar{C}_X^2, \quad
c_{3,0} = -\frac{1}{2} \bar{\lambda}_{\rm HX} \bar{C}_X, \quad
c_{2,0} = 2 (C^2 + \tilde{C}^2) + \frac{1}{4} \bar{\lambda}_{\rm HX}^2, \quad
M = \Lambda.
\label{eq:matching_coefficients}
\end{equation}
The distinct-component channel from $\tilde{\mathcal O}_X$ is not contained
in the diagonal $c_{3,0}$ above.  After its channel multiplicity is included,
the coefficient entering the thermally summed generic-rate notation is
\begin{equation}
c_{3,0}^{\rm eff}
=\frac{\bar{\tilde C}_X^2}{12}
-\frac{1}{2}\bar\lambda_{HX}\bar C_X,
\label{eq:c30_eff}
\end{equation}
which directly gives the two terms in $F_5$ below.
Therefore, in the limit of $T_\mathrm{reh} \gg m_h$, the DM yield is given by
\begin{align}
Y_X(m_h) \simeq 2\times \frac{g_\phi}{\sqrt{g_*(T_{\rm reh})}}
\left(F_7 T^7_\mathrm{reh} + F_5 T^5_\mathrm{reh} + F_3 T^3_\mathrm{reh}\right),
\label{eq:yield_general}
\end{align}
where
\begin{align}
F_7 &= \frac{32 \sqrt{2/5} \, \pi^6 M_{\rm pl} \, \bar{C}_X^2}{9261 \, \Lambda^8},\\
F_5 &= \frac{18 \sqrt{2/5} \, M_{\rm pl} \, \zeta(5)^2}
{\pi^6 \, \Lambda^6}
\left(\bar{\tilde C}_X^2-6\bar{\lambda}_{HX}\bar{C}_X\right),
\label{eq:F_5}\\
F_3 &= \frac{\pi^2 M_{\rm pl}}
{540 \sqrt{10}\Lambda^4} 
\left[ 2 \left(C^2 + \tilde{C}^2\right) + \frac{1}{4} \bar{\lambda}_{HX}^2
\right]. 
\end{align}
The DM relic density at present is determined as
\begin{align}
\Omega_X h^2 
&= 1.6 \times 10^8 \, 
\left(\frac{m_X}{1~\mathrm{GeV}}\right)
\left[\frac{g_{*s}(T_0)}{g_{*s}(T_{\rm reh})}\right]
Y_X(m_h).
\label{eq:DM_relic_general}
\end{align}
As explained in Sec.~\ref{sec:dm_relic_setup}, the lower bound on $\Lambda/T_{\rm reh}$
follows from the out-of-equilibrium condition and depends on the reheating temperature and the Wilson coefficients;
Eq.~\eqref{eq:cutoff_fimp} gives the order-of-magnitude estimate for order-unity field-strength coefficients.
Since we allow general hierarchies among the Wilson coefficients, we retain the $F_3$, $F_5$ and $F_7$ terms in Eq.~\eqref{eq:yield_general}.
We consider two representative cases:
\begin{description}
    \item[Case 1.] The coefficients $C$, $\tilde{C}$,
    $\bar{\lambda}_{HX}$, $\bar{C}_X$, and $\bar{\tilde C}_X$ are at
    most of similar order. From Eq.~\eqref{eq:coeff_cond3}, we focus on
    values of order 0.1. The terms proportional to $F_5$ and $F_7$ are
    suppressed by additional powers of $T_{\rm reh}/\Lambda$.
    \item[Case 2.] Motivated by the tree-versus-loop matching pattern described
    above, the coefficient $\bar{C}_X$ is much larger than the others.
    Specifically, we take the coefficients to be
    $\bar{C}_X=0.1 \gg
    \{ C, \tilde{C}, \bar{\lambda}_{HX},\bar{\tilde C}_X \}$,
    with $C=\tilde{C}$. We set $\bar{\tilde C}_X=0$ in the plots to
    isolate the sign-dependent interference between
    $\bar\lambda_{HX}$ and $\bar C_X$; a nonzero
    $\bar{\tilde C}_X$ adds the positive term shown in Eq.~\eqref{eq:F_5}.
\end{description}
In the following, we will discuss these two cases in turn.
\paragraph{Case 1.}
For Case 1, the $F_7$ and $F_5$ terms can be neglected.
By replacing $C^2 + \tilde{C}^2$ with $C^2 + \tilde{C}^2 + \bar{\lambda}_{\rm HX}^2/8$ in Eq.~\eqref{eq:DM_relic1}, the DM relic density is given by
\begin{align}
\Omega_X h^2 &\simeq 0.12
\left(\frac{m_X}{100~\mathrm{GeV}}\right)
\left(\frac{T_{\mathrm{reh}}}{10^{10}~\mathrm{GeV}}\right)^3
\left(\frac{10^{14}~\mathrm{GeV}}{\Lambda}\right)^4
\times 1.1 \left(C^2 + \tilde{C}^2 + \frac{1}{8}\bar{\lambda}_{HX}^2 \right).
\label{eq:DM_relic_stu1}
\end{align}
For $C^2 + \tilde{C}^2 + \bar{\lambda}_{HX}^2/8 \simeq 0.01$, corresponding to Case 1, the DM mass required to reproduce the observed relic density is increased by approximately two orders of magnitude compared to the case in which only the gauge-invariant dimension-6 operators with $\mathcal{O}(1)$ coefficients are present.
Alternatively, choosing a lower cutoff scale provides another viable option.

In Case 1, the relic abundance at fixed dark matter mass and Wilson coefficients scales as $T_{\mathrm{reh}}^3/\Lambda^4$. Whether reducing both scales enhances the abundance therefore depends on their relative variation. Any choice of scales must also satisfy the out-of-equilibrium condition and our characteristic-energy criterion.

\paragraph{Case 2.}
For Case 2, in addition to the $F_3$ term, the $F_7$ and $F_5$ terms can also contribute to DM production.
Since $C$ and $\tilde{C}$ enter the $F_3$ term in a similar manner, we take $C = \tilde{C}$.
Although $\bar{\lambda}_{HX}$ also contributes in a similar way, it additionally enters the $F_5$ term through an interference effect with $\bar{C}_X$.
Therefore, we do not assume $\bar{\lambda}_{HX}$ to have the same value as $C$ and $\tilde{C}$.

\begin{figure}[!t]
\begin{center}
\includegraphics[width=0.40\textwidth,clip]{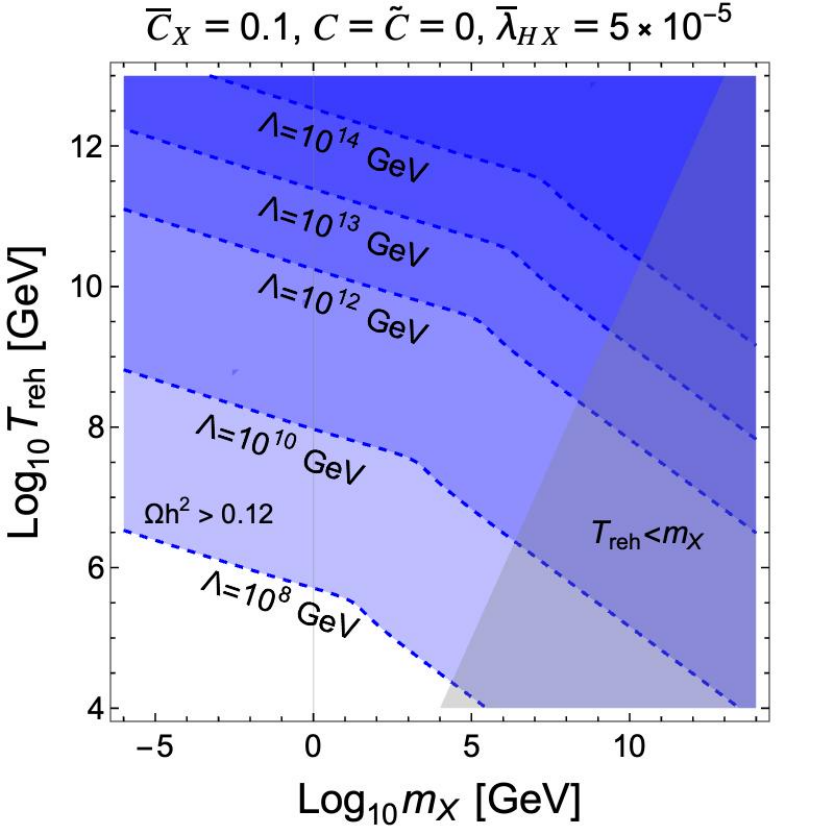}\,\,
\includegraphics[width=0.40\textwidth,clip]{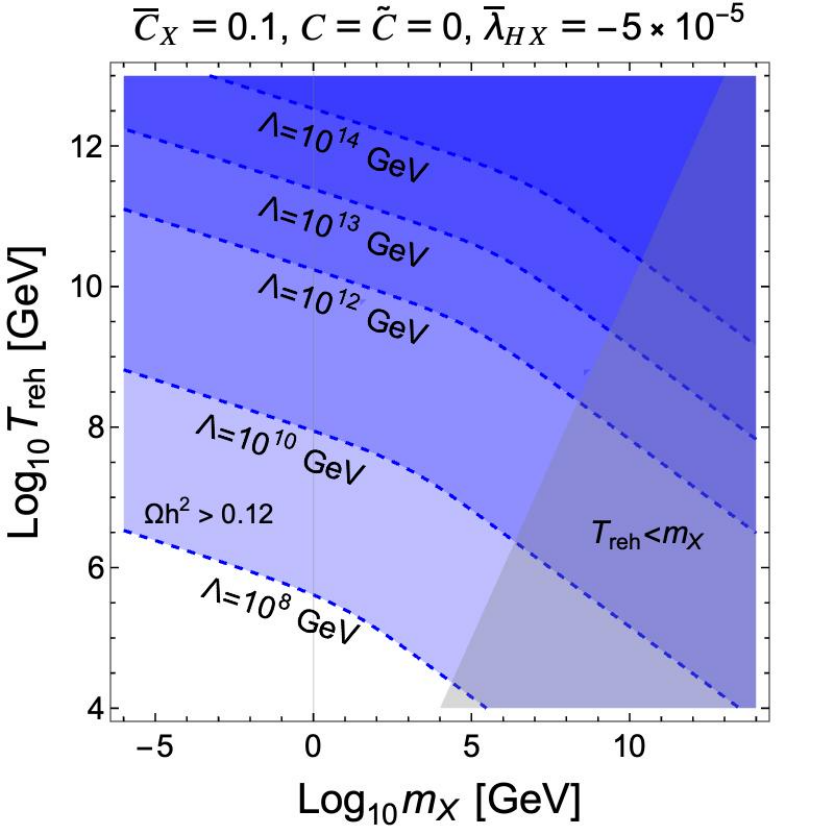}
\end{center}
\caption{
Same as Figure~\ref{fig:RelicD}, but with a nonzero coupling $\bar{C}_X = 0.1$, whereas the gauge-invariant dimension-6 operators are set to zero, $C = \tilde{C} = 0$. In the left and right panels, 
we take $\bar{\lambda}_{HX}=5\times10^{-5}$ and $\bar{\lambda}_{HX}=-5\times10^{-5}$, respectively.
We also set $\bar{\tilde C}_X=0$ to isolate the interference term.
The cutoff scale is chosen as $\Lambda = 10^{8,10,12,13,14}~\mathrm{GeV}$.
Segments that fail the characteristic-energy criterion
$10.9T_{\rm reh}<\Lambda$ are retained only to display the analytic scaling
and are not claimed to satisfy our characteristic-energy criterion; see Sec.~\ref{sec:cosmology}.
%The grey regions indicate $T_{\mathrm{reh}} < m_X$, 
%where DM production is Boltzmann suppressed.  
%The DM relic density is overproduced in the blue region 
%($\Omega_X h^2 > 0.12$), whereas the observed value 
%is realized along the boundary.
}
\label{fig:RelicD-stu}
\end{figure}

In Figure~\ref{fig:RelicD-stu}, we show the parameter space in the $m_X$-$T_{\rm reh}$ plane that reproduces the observed DM relic density.
For the cutoff scale, we take $\Lambda = 10^{8,10,12,13,14}~\mathrm{GeV}$.
We set the gauge-invariant dimension-6 operators and the parity-odd
Stueckelberg coefficient to zero, $C = \tilde{C} =\bar{\tilde C}_X= 0$,
in order to see the effects of the $F_5$ interference term and the $F_7$
term from $\mathcal O_X$.
We take $\bar{\lambda}_{HX} = 5 \times 10^{-5}$ in the left panel and $\bar{\lambda}_{HX} = -5 \times 10^{-5}$ in the right panel, respectively.
Throughout this figure, we fix $\bar{C}_X = 0.1$.
For a given cutoff scale, the $F_7$ term dominates at higher reheating temperatures, whereas the $F_3$ term becomes dominant at lower reheating temperatures in both panels.
By comparing the left and right panels, one can also observe mild destructive (constructive) interference effects for positive (negative) values of $\bar{\lambda}_{HX}$.
Note that, since the interference term $F_5$ in Eq.~\eqref{eq:F_5} is proportional to $\bar{\lambda}_{HX}\bar{C}_X$, the same effect can be obtained by flipping the sign of $\bar{C}_X$ instead of $\bar{\lambda}_{HX}$.
\begin{figure}[!t]
\begin{center}
\includegraphics[width=0.45\textwidth,clip]{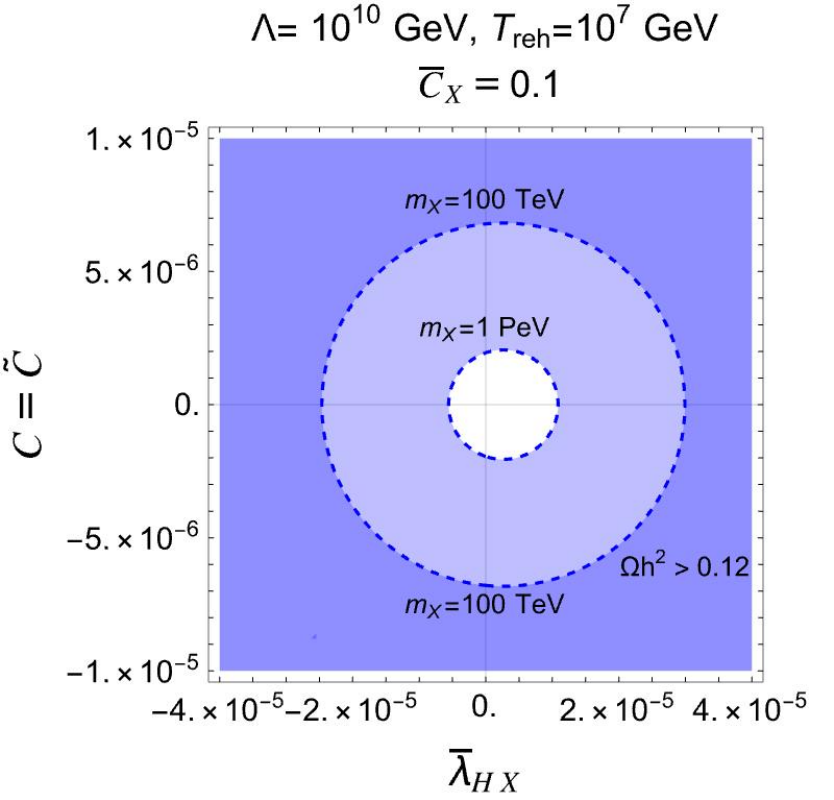} \,\,
\includegraphics[width=0.45\textwidth,clip]{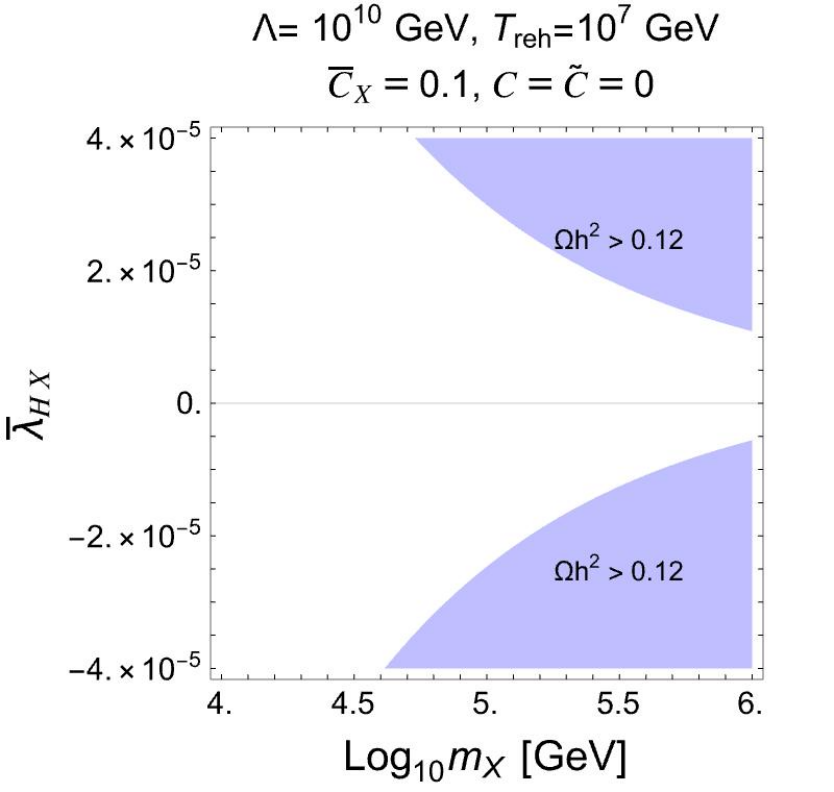}
\end{center}
\caption{Left: Parameter space in the $\bar{\lambda}_{HX}$-$C(=\tilde{C})$ 
plane. We take $m_X = 100~\mathrm{TeV}$ and $1~\mathrm{PeV}$.  
Right: Parameter space 
in the $m_X$-$\bar{\lambda}_{HX}$ plane. 
In the right panel, the gauge-invariant dimension-6 operators are set to zero, $C = \tilde{C} = 0$.
In all plots, we fix $\bar{C}_X=0.1$, $\bar{\tilde C}_X=0$,
$\Lambda = 10^{10}~\mathrm{GeV}$, and $T_{\mathrm{reh}} = 10^{7}~\mathrm{GeV}$.}
\label{fig:RelicD2-stu}
\end{figure}

In Figure~\ref{fig:RelicD2-stu}, we present the parameter space in the 
$\bar{\lambda}_{HX}$-$C(=\tilde{C})$ plane (left) and in the $m_X$-$\bar{\lambda}_{HX}$ plane (right) 
that reproduces the observed DM relic density.  
In the right panel, the gauge-invariant dimension-6 operators are set to zero, i.e., $C = \tilde{C} = 0$.
Here, we fix $\Lambda = 10^{10}$~GeV and $T_{\mathrm{reh}} = 10^{7}$~GeV. 
In the left panel, for $m_X = 1~\mathrm{PeV}$ and $m_X = 100~\mathrm{TeV}$, the boundaries corresponding to $\Omega_X h^2 = 0.12$ are obtained for small coefficients of order $10^{-6}$-$10^{-5}$.
The boundary curves in the right panel indicate that $|\bar{\lambda}_{HX}|$ decreases as $m_X$ increases.
From both panels, we observe destructive (constructive) interference for positive (negative) values of $\bar{\lambda}_{HX}$.

Finally, we comment on the freeze-in DM production.
One may wonder whether processes schematically of 
the form $XX \to \phi\phi$ (i.e., the inverse processes) 
could spoil DM production.
The Higgs components $\phi_i$ are maintained in thermal equilibrium by SM
gauge, Yukawa, and scalar interactions. This is precisely what justifies the
Bose--Einstein distributions in Eq.~\eqref{eq:bose_d}. The portal condition
$n_\phi^{\rm eq}\langle\sigma v\rangle_{\phi\phi\to XX}<H$ does not state
that the Higgs bath is out of equilibrium; it states that the portal fails to
establish mutual chemical equilibrium between the SM bath and $X$.

The inverse process $XX\to\phi\phi$ is neglected at leading order in the
freeze-in approximation. Its collision term is quadratic in the nonthermal
phase-space distribution of $X$, whereas the forward production term is
sourced by the thermal Higgs bath. The small values of the portal-rate
diagnostic in Appendix~\ref{app:consistency} provide an additional consistency
check that the portal interactions do not establish chemical equilibrium.
As a result, the Boltzmann equation for the number density of $X$ can be written as in Eq.~\eqref{eq:boltzmann}.

%%%%%%%%%%%%%
%%%%%%%%%%%%%
% 
%%%%%%%%%%%%%
%%%%%%%%%%%%%

\subsection{Cosmological Scope and Additional Constraints}
\label{sec:cosmology}

Our calculation assumes that a radiation-dominated epoch begins at
$T_{\rm reh}$ and that the initial abundance of $X$ is negligible. In an
extended reheating history, the maximum temperature $T_{\max}$ can exceed
$T_{\rm reh}$ and production before radiation domination can alter a UV
freeze-in abundance~\cite{Bernal:2019mhf}. This qualification is especially
important for the $T_{\rm reh}^7$ term. Thus, $T_{\rm reh}$ in our numerical
analysis should be understood as the highest temperature included in the
sudden-reheating, radiation-dominated calculation. A specified inflaton sector
would be required to translate our contours to a noninstantaneous reheating
history.

The range $T_{\rm reh}=10^6$--$10^{14}$~GeV is not subject to a universal
upper bound in the minimal SM plus $X$ cosmology. Possible gravitino, modulus,
or topological-defect constraints depend on additional UV degrees of freedom.
Likewise, $\Lambda\simeq10^{16}$~GeV may be interpreted as a high matching
scale, comparable to conventional grand-unification scales and below
$M_{\rm pl}$, but it is not by itself a solution to the electroweak hierarchy
problem. A generic heavy state coupled to the Higgs can give a threshold
correction of order $g_H^2\Lambda^2/(16\pi^2)$; a UV completion at this scale
therefore requires a protective structure, small couplings, or tuning.

EFT validity must apply to the center-of-mass energies sampled by the thermal
integral, not only to $T_{\rm reh}$. For the $s^2$--$s^4$ terms, the thermal
moments give characteristic center-of-mass energies of approximately
$6.6T$--$10.9T$. We consequently require the dominant thermally sampled
energies to satisfy our characteristic-energy criterion based on
$\Lambda_{\rm eff}$ and apply this condition point by point. This
characteristic-energy prescription does not by itself bound the fraction of
the thermal collision integral arising from energies above $\Lambda_{\rm eff}$.
It is comfortably satisfied for the field-strength examples
and for the Case~2 curves with $\Lambda=10^8$ and $10^{10}$~GeV above their
structure-formation reference masses.  For the $s^4$-dominated Case~2 curves,
the equality $10.9T_{\rm reh}=\Lambda$ occurs at
$m_X\simeq9.47\ (9.43)$~keV for $\Lambda=10^{12}$~GeV,
$94.7\ (94.3)$~keV for $\Lambda=10^{13}$~GeV, and
$947\ (943)$~keV for $\Lambda=10^{14}$~GeV, where the values outside
(inside) parentheses correspond to
$\bar\lambda_{HX}=+5\times10^{-5}$ ($-5\times10^{-5}$).
Only the portions above these masses are retained as satisfying our
characteristic-energy criterion.
The lower-mass portions remain in Fig.~\ref{fig:RelicD-stu} solely to show
the analytic relic-density scaling.

The exact $Z_2$ symmetry makes $X$ stable, while the freeze-in condition makes
late annihilation or energy injection negligible. When $X$ is still
relativistic during big bang nucleosynthesis (BBN), entropy dilution from
$g_{*s}=106.75$ gives the estimate
\begin{align}
\Delta N_{\rm eff}^{\rm BBN}\simeq
6.4\times10^{-4}
\left(\frac{\langle p/T\rangle_{\rm prod}}{3.83}\right)
\left(\frac{10~{\rm keV}}{m_X}\right)
\left(\frac{\Omega_Xh^2}{0.12}\right).
\label{eq:dneff}
\end{align}
For the momentum moments and conservative masses in
Appendix~\ref{app:lyman}, this is at most a few $10^{-3}$, well below the
present sensitivity to extra radiation. The retained dark matter is
nonrelativistic long before recombination, so the CMB expansion constraint is
also weaker than the small-scale-structure constraint. We therefore find
Lyman-$\alpha$ structure formation to be the leading cosmological limitation
on the low-mass region, consistently with the general freeze-in analysis of
Ref.~\cite{Decant:2021mhj}. Inflationary, inflaton-decay, and purely
gravitational production of $X$ are model dependent and have been assumed
negligible; they can become relevant if all portal coefficients are extremely
small.

\section{Summary}\label{sec:summary}

We studied UV freeze-in of a $Z_2$-odd dark photon whose mass is generated by
the Stueckelberg mechanism. Besides the parity-even and parity-odd
field-strength portals, the gauge-invariant Stueckelberg parent theory relates
the unitary-gauge couplings of longitudinal vectors to powers of
$m_X/\Lambda$. This relation, rather than the replacement of a Proca mass
alone, is the central new ingredient of the analysis.

We made explicit that placing all interactions in one EFT with
$\Lambda_X,\Lambda_X^{\rm dim-6},
\Lambda_X^{\widetilde{\mathcal O}}>\Lambda$ is a conservative
single-cutoff assumption, not a consequence that every UV completion must
satisfy. Under this power counting, order-one $C$ and $\tilde C$ make the
field-strength portals dominant. In a multiscale theory, the general
condition is instead that the characteristic production energies lie below
$\Lambda_{\rm eff}$, and the Stueckelberg terms can be larger.

When the coefficients have a tree-versus-loop hierarchy, the longitudinal
operators generate yields proportional to $T_{\rm reh}^3$,
$T_{\rm reh}^5$, and $T_{\rm reh}^7$, including a sign-sensitive
$\bar\lambda_{HX}\bar C_X$ interference term. We supplied a schematic heavy
mediator matching pattern that motivates this hierarchy while stressing that
Case~2 is a benchmark rather than a generic prediction. The
parity-odd operator $\tilde{\mathcal O}_X$ produces
longitudinal--transverse final states and a positive contribution
proportional to $\bar{\tilde C}_X^2T_{\rm reh}^5$.

The observed abundance can be reproduced over a broad mass range. For
example, $m_X\simeq10$~TeV is compatible with
$\Lambda\simeq10^{16}$~GeV and $T_{\rm reh}\simeq10^{12}$~GeV for
order-one field-strength coefficients, while the hierarchical benchmark
allows masses around $100$~TeV for $\Lambda\simeq10^{10}$~GeV and
$T_{\rm reh}\simeq10^7$~GeV. We explicitly checked the portal rate and find
${\cal R}_{\rm FI}(T)<3.4\times10^{-2}$ throughout
$m_h\leq T\leq T_{\rm reh}$ along the displayed relic-density
contours at and above the operator-matched conservative
structure-formation reference scales.  We also identified and excluded from
the interpretation based on our characteristic-energy criterion
the high-cutoff, low-mass portions for
which the characteristic thermal energy exceeds the cutoff.
Thermal Higgs masses change the UV rates by less than about six percent in
our conservative estimate.
For the nonthermal momentum distribution considered here, we use an
operator-dependent estimate of the structure-formation reference masses.
Indicative lower masses are $6$--$9$~keV with a conservative
Lyman-$\alpha$ calibration and $23$--$35$~keV with a stringent one; a
momentum-dependent Boltzmann and transfer-function calculation is required
for a sharp exclusion. BBN and CMB limits are subdominant in this range.
Noninstantaneous reheating and model-dependent production before radiation
domination remain important directions for future work.

An additional model-building assumption underlying the present analysis is that
the dark $Z_2$ parity is exact. We do not specify the origin of this discrete
symmetry. Its exactness both guarantees the stability of $X$ and forbids kinetic
mixing with $U(1)_Y$; accordingly, the phenomenology studied here relies on
treating this $Z_2$ as an exact symmetry.

\paragraph*{Acknowledgments}
We have used the package TikZ-Feynman~\cite{Ellis:2016jkw} to draw the Feynman diagrams.
K.~Y. would like to thank Takashi Shimomura for valuable discussions.
S.~T. is supported by JSPS KAKENHI Grant No.~22K03642 and 
Waseda University Special Research Projects (No.~2025C-488 and No.~2025R-028).
The work of K.~Y. was supported in part by JSPS KAKENHI Grant No.~JP24K17040.

\paragraph*{Data Availability}
There are no publicly available research data or software supporting this manuscript.
Requests for further information or data should be sent to the authors.

\appendix

\section{Parity-Odd Stueckelberg Operator}
\label{app:tildeOX}

We derive here the contribution of
$\tilde{\mathcal O}_X=i(H^\dagger\overleftrightarrow{D^\mu}H)
\tilde X_{\mu\nu}X^\nu$. In terms of the real Higgs components, the
antisymmetric current connects $\phi_1$ with $\phi_2$ and $\phi_3$ with
$\phi_4$. For $\phi_1(p_1)\phi_2(p_2)\to
X(k_1,\epsilon_1)X(k_2,\epsilon_2)$, the amplitude, up to an irrelevant
overall phase, is
\begin{align}
{\cal M}_{\tilde{\mathcal O}_X}
=\frac{\tilde C_X}{\Lambda^2}(p_1-p_2)^\mu
\epsilon_{\mu\nu\rho\sigma}
\left[k_1^\rho\epsilon_1^\sigma\epsilon_2^\nu
+k_2^\rho\epsilon_2^\sigma\epsilon_1^\nu\right].
\label{eq:tilde_amp}
\end{align}
The Levi--Civita tensor gives the parity-odd tensor structure. A separate CP
assignment depends on the charge-conjugation convention adopted for the Higgs
current and the dark vector and is not needed for the production calculation.
In the COM frame its contraction
vanishes for longitudinal--longitudinal and transverse--transverse
polarizations; only the longitudinal--transverse and
transverse--longitudinal states survive. Summing over final polarizations gives
\begin{align}
\sum_{\rm pol}|{\cal M}_{\tilde{\mathcal O}_X}|^2
=\frac{\tilde C_X^2}{\Lambda^4}
\frac{s(s-4m_X^2)^2}{2m_X^2}\sin^2\theta ,
\label{eq:tilde_amp_sq}
\end{align}
before the $1/2!$ symmetry factor for the identical final vectors,
where $\theta$ is the scattering angle in the COM frame.

For equal unitary-gauge coefficients, Eq.~\eqref{eq:tilde_amp_sq} is
suppressed by $m_X^2/s$ relative to $\mathcal O_X$. After angular integration
and accounting for the two distinct Higgs pairs versus the four diagonal
species, the ratio of the fixed-$s$, channel- and angle-summed
event-density kernels is
\begin{align}
\frac{{\rm d}\gamma_{\tilde{\mathcal O}_X}/{\rm d}s}
{{\rm d}\gamma_{\mathcal O_X}/{\rm d}s}
\simeq\frac{4}{3}\frac{\tilde C_X^2}{C_X^2}\frac{m_X^2}{s}.
\label{eq:tilde_equal_ratio}
\end{align}
This does not justify dropping the operator under natural Stueckelberg power
counting. Indeed,
$\tilde C_X=\bar{\tilde C}_Xm_X/\Lambda$, whereas
$C_X=2\bar C_Xm_X^2/\Lambda^2$. For comparable barred coefficients,
$|{\cal M}_{\tilde{\mathcal O}_X}|^2\sim
\bar{\tilde C}_X^2s^3/\Lambda^6$ and
$|{\cal M}_{\mathcal O_X}|^2\sim\bar C_X^2s^4/\Lambda^8$.
The former can therefore be the larger contribution within the EFT regime.

Performing the Bose--Einstein phase-space integral and summing the two
distinct Higgs pairs yields
\begin{align}
\gamma_{\tilde{\mathcal O}_X}(T)
=\frac{24\zeta(5)^2}{\pi^5}
\frac{\bar{\tilde C}_X^2T^{10}}{\Lambda^6}.
\label{eq:tilde_rate}
\end{align}
It does not interfere with the diagonal $\phi_i\phi_i$ amplitudes. Integration
of Eq.~\eqref{eq:boltzmann_yield} gives the positive
$\bar{\tilde C}_X^2$ part of $F_5$ displayed in Eq.~\eqref{eq:F_5}.

\section{Freeze-In and Thermal Mass Checks}
\label{app:consistency}

Let $\gamma$ be the event density in Eq.~\eqref{eq:rate}. Since each event
creates two vectors, the appropriate rate diagnostic is
\begin{align}
{\cal R}_{\rm FI}(T)
\equiv\frac{2\gamma(T)}{n_\phi^{\rm eq}(T)H(T)}
=\frac{2\pi^2}{\zeta(3)\sqrt{g_*}}
\left[7F_7T^7+5F_5T^5+3F_3T^3\right].
\label{eq:rfi_exact}
\end{align}
For the field-strength portals alone this reduces to
\begin{align}
{\cal R}_{\rm FI}(T)=
\frac{2\pi^4}{2700\zeta(3)}
\sqrt{\frac{90}{g_*}}\,
(C^2+\tilde C^2)\frac{M_{\rm pl}T^3}{\Lambda^4}.
\label{eq:rfi_gauge}
\end{align}
Table~\ref{tab:rfi} gives representative points on the
$\Omega_Xh^2=0.12$ boundaries, evaluated at the largest temperature in each
benchmark. The sign-dependent entries use the full expression in
Eq.~\eqref{eq:rfi_exact}.
\begin{table}[t]
\centering
\small
\setlength{\tabcolsep}{3pt}
\renewcommand{\arraystretch}{1.15}
\begin{tabular}{c c c c c c}
\hline
benchmark & $m_X$ & $T_{\rm reh}$ [GeV] & $\Lambda$ [GeV] &
coefficients & ${\cal R}_{\rm FI}$\\
\hline
field strength & $4.70$ TeV & $10^{12}$ & $10^{16}$ &
$C=\tilde C=1$ & $2.68\times10^{-11}$\\
field strength & $0.470$ GeV & $10^{8}$ & $10^{12}$ &
$C=\tilde C=1$ & $2.68\times10^{-7}$\\
Case 2 & $33.4$ TeV & $10^{7}$ & $10^{10}$ &
$\begin{gathered}\bar C_X=0.1,\\
\bar\lambda_{HX}=+5\times10^{-5}\end{gathered}$ &
$3.51\times10^{-12}$\\
Case 2 & $27.0$ TeV & $10^{7}$ & $10^{10}$ &
$\begin{gathered}\bar C_X=0.1,\\
\bar\lambda_{HX}=-5\times10^{-5}\end{gathered}$ &
$4.99\times10^{-12}$\\
\hline
\end{tabular}
\caption{Freeze-in rate diagnostic for representative relic-density points.
All unlisted coefficients are zero. We use $g_*=106.75$.}
\label{tab:rfi}
\end{table}
The contour-wide bound can be obtained analytically rather than from a
parameter scan.  For a single monomial contribution
$Y_X\propto T_{\rm reh}^{\,n}$, Eqs.~\eqref{eq:boltzmann_yield} and
\eqref{eq:rfi_exact} imply, at the upper end of the integration interval,
\begin{align}
{\cal R}_{\rm FI}(T_{\rm reh})
=\frac{\pi^2n}{g_\phi\zeta(3)}Y_X,\qquad
Y_X=\frac{0.12}
{1.6\times10^8(m_X/{\rm GeV})
[g_{*s}(T_0)/g_{*s}(T_{\rm reh})]} .
\label{eq:rfi_contour}
\end{align}
At the operator-matched conservative endpoints, the
$(n,m_X)=(3,5.9~{\rm keV})$, $(5,7.4~{\rm keV})$, and
$(7,8.8~{\rm keV})$ cases instead give $0.0214$, $0.0284$, and $0.0334$,
respectively.  Direct evaluation of the complete Case~2 polynomials at the
$s^4$-matched reference mass $m_X=8.8$~keV gives maximal values of
$0.0337$ and $0.0334$ over the displayed cutoff choices for the positive and
negative interference signs, respectively; both rates increase monotonically
over the relevant temperature interval.
Hence
${\cal R}_{\rm FI}(T)<3.4\times10^{-2}$ throughout
$m_h\leq T\leq T_{\rm reh}$ on the displayed relic-density contours at and above the operator-matched
conservative structure-formation reference masses.  The blue overabundant
regions are not part of this claim; sufficiently far inside them the
freeze-in approximation can fail.

For completeness, we quantify the thermal Higgs mass. Write
$m_{H,\mathrm{th}}^2=c_HT^2$ and $q=m_{H,\mathrm{th}}/T$. In a
Maxwell--Boltzmann estimate, the phase-space suppression for an amplitude
squared proportional to $s^r$ is
\begin{align}
{\cal S}_r(q)=
\frac{\displaystyle\int_{2q}^{\infty}{\rm d}x\,
x^{2r+1}\sqrt{x^2-4q^2}\,K_1(x)}
{\displaystyle\int_0^{\infty}{\rm d}x\,x^{2r+2}K_1(x)}.
\label{eq:thermal_supp}
\end{align}
At the conservative value $c_H=0.40$, and hence $q=0.63$, this gives
${\cal S}_2=0.965$, ${\cal S}_3=0.983$, and ${\cal S}_4=0.990$.
Repeating the calculation with the Bose--Einstein distributions used in the
main text gives $0.944$, $0.978$, and $0.988$, respectively. Including the
mass-dependent derivative vertices, all the retained $F_3$, $F_5$, and $F_7$
rates change by no more than about six percent. We therefore retain the
massless analytic rates with this small normalization uncertainty.

\section{Momentum Moments and Structure Formation}
\label{app:lyman}

The freeze-in population is nonthermal, so a thermal-WDM mass bound
cannot be applied directly.
Reference~\cite{Decant:2021mhj} finds a stringent $15$~keV limit for its standard
pure-freeze-in distribution and discusses a conservative calibration near
$3.5$~keV. Our contact processes are more UV weighted. We evaluated the
phase-space moments of the daughter momentum for each high-energy dependence
$|{\cal M}|^2\propto s^r$.  We define $q\equiv p_X/T$. The massless
Bose--Einstein phase-space integrals give
\begin{align}
\langle q\rangle_r
&=(r+2)\frac{\zeta(r+3)}{\zeta(r+2)},\nonumber\\
\langle q^2\rangle_r
&=\frac{(r+2)(r+3)}{3}
\left[
2\frac{\zeta(r+4)}{\zeta(r+2)}
+\left(\frac{\zeta(r+3)}{\zeta(r+2)}\right)^2
\right].
\label{eq:moment_q}
\end{align}
Following the rms-velocity mapping convention of
Ref.~\cite{Decant:2021mhj}, the corresponding reference mass is
\begin{align}
m_X^{\rm ref}\simeq&
1.75~{\rm keV}\,\sqrt{\langle q^2\rangle_r}
\left(\frac{3.91}{106.75}\right)^{1/3}
\left(\frac{m_{\rm WDM}^{\rm Ly\alpha}}{1~{\rm keV}}\right)^{4/3}.
\label{eq:lyman_mapping}
\end{align}
We use $m_{\rm WDM}^{\rm Ly\alpha}=1.9$~keV for the conservative thermal-WDM
calibration and $5.3$~keV for the stringent one.  The resulting rms momenta
and reference masses are shown in Table~\ref{tab:lyman}.
\begin{table}[H]
\centering
\begin{tabular}{c c c c}
\hline
dominant power & $\sqrt{\langle(p_X/T)^2\rangle}$ &
conservative reference [keV] & stringent reference [keV]\\
\hline
$s^2$ & $4.32$ & $5.9$ & $23.2$\\
$s^3$ & $5.39$ & $7.4$ & $29.0$\\
$s^4$ & $6.43$ & $8.8$ & $34.6$\\
\hline
\end{tabular}
\caption{Momentum moments and indicative mass scales obtained by mapping the
rms velocity to the conservative and stringent calibrations of
Ref.~\cite{Decant:2021mhj}.}
\label{tab:lyman}
\end{table}
Equation~\eqref{eq:moment_q} assumes that, after extracting the factor $s^r$,
the squared amplitude is independent of the COM scattering angle. The retained
operator $\widetilde{\mathcal O}_X$ instead has
$|\mathcal M|^2\propto s^3\sin^2\theta$. Repeating the angular average gives
\begin{align}
\left\langle q^2\right\rangle_{r,\sin^2\theta}
&=\frac{(r+2)(r+3)}{5}\left[
3\frac{\zeta(r+4)}{\zeta(r+2)}
+2\left(\frac{\zeta(r+3)}{\zeta(r+2)}\right)^2
\right].
\label{eq:moment_q_sin2}
\end{align}
For $r=3$, this gives
$\sqrt{\langle q^2\rangle}=5.3903$ instead of $5.3921$, a relative shift of
about $0.034\%$. This is negligible at the accuracy of the indicative mapping
in Table~\ref{tab:lyman}, whose rounded value $5.39$ is therefore unchanged,
but we record it for completeness.
The corresponding mean momenta are $3.83T$, $4.91T$, and $5.95T$. Entropy
release between production and late times is included using
$g_{*s}(T_{\rm prod})=106.75$. Because several powers can contribute and
interfere, the entries in Table~\ref{tab:lyman} are reference scales rather
than exclusion limits. A robust bound requires the unintegrated Boltzmann
equation for the complete distribution followed by a
linear-transfer-function calculation. Accordingly, portions of the plotted
abundance contours below these scales are retained only to show analytic
parameter dependences and are not claimed as cosmologically viable.

%%%%%%%%%%%%%%%%%%%%%%%%%%%
%%%%%%%%%%%%%%%%%%%%%%%%%%%

\end{document}